\documentclass{article}
\usepackage{pdfpages}
\usepackage{xcolor}
\usepackage{graphicx}
\usepackage{cite}
\usepackage{caption}
\usepackage{subcaption}
\usepackage{caption}
\usepackage{authblk}
\usepackage{amsmath}
\usepackage[top=5cm, bottom=5cm, left=1cm, right=1cm, headsep=1.5cm, heightrounded=true]{geometry}
\usepackage{fancyhdr}
\usepackage[labelfont=bf, labelsep=colon, format=plain]{caption}

\usepackage{csquotes}
\usepackage{xcolor}

\usepackage{nameref}

\usepackage{newtxtext} 
\usepackage{lmodern}

\usepackage{setspace}

\usepackage[colorlinks=true,linkcolor=blue,citecolor=blue,urlcolor=blue]{hyperref}

\usepackage{float}

\title{ Reconstruction of the Quintessence Scalar Field Potential Using Gaussian Processes }
\author{Redouane El Ouardi$^{1}$\thanks{ \texttt{redouane.elouardi.d23@ump.ac.ma}}}
\author{Amine Bouali$^{1,2,3}$\thanks{ \texttt{a1.bouali@ump.ac.ma}}}
\author{Ahmed Errahmani$^{1,2}$\thanks{ \texttt{ahmederrahmani1@yahoo.fr}}}
\author{Ryan E. Keeley$^{4}$\thanks{ \texttt{rkeeley@uci.edu}}}

\author{Arman Shafieloo$^{5}$\thanks{ \texttt{shafieloo@kasi.re.kr}}}
\author{Taoufik Ouali$^{1,2}$\thanks{ \texttt{t.ouali@ump.ac.ma}}}
\affil {\textit{$^{1}$Laboratory of Physics of Matter and Radiation, }\\
		\centering \textit{ Mohammed I University, BP 717,  Morocco}\\
		\textit{$^{2}$Astrophysical and Cosmological Center, Faculty of Sciences, BP 717, Morocco}\\
		\textit{$^{3}$Higher School of Education and Training, Mohammed I University, BP 717, Morocco}\\
        \textit{$^{4}$Department of Physics and Astronomy, University of California, Irvine, California 92697-4575, USA}\\
        \textit{$^{5}$Korea Astronomy and Space Science Institute, 776, Daedeokdae-ro, Yuseong-gu, Daejeon
34055, Republic of Korea}}

\begin{document}
\maketitle

\begin{abstract}\

Recent cosmological observations, including the latest Dark Energy Spectroscopic Instrument (DESI) data releases DR1 and DR2, have renewed interest in the possibility that dark energy may exhibit
dynamical behavior rather than being a strict cosmological constant. In this work, we perform
a fully model-independent reconstruction of the quintessence scalar field potential using
Gaussian Process regression and current Hubble measurements. Instead of assuming a specific functional form for the scalar field potential, we reconstruct the quintessence potential and the corresponding kinetic energy directly from observational data. Our analysis is based on Hubble parameter measurements obtained from cosmic chronometers and the latest high-precision DESI DR2 baryon acoustic oscillation (BAO) data, together with Type Ia supernova data from the Pantheon+ compilation. Gaussian Processes provide a nonparametric
and model-independent framework that allows the data to guide the reconstruction. We employ two covariance functions, namely the squared exponential and the Matern ($\nu = 9/2$) kernels, in order to assess the sensitivity of the reconstruction to the kernel
choice. We further explore the impact of background cosmological assumptions by considering
different priors on the matter density and spatial curvature. Finally, we compare the reconstructed scalar field potential with two theoretically motivated
benchmark models: a power law potential and an exponential potential. We find that both models
remain consistent with the reconstructed potential within the inferred confidence intervals.
\end{abstract}
\textbf{Keywords:} dark energy; quintessence; Gaussian Processes; DESI DR2; cosmological reconstruction

\section{ Introduction}\ 

The observational evidence for the late-time acceleration of the Universe emerged in the late 1990s, following the independent analyses of distant Type Ia supernovae by Riess et al. \cite{SupernovaSearchTeam:1998fmf} and Perlmutter et al. \cite{perlmutter1999measurements}. These results established accelerated cosmic expansion as a robust empirical fact and have since become a key feature of modern cosmology. Identifying the physical mechanism, responsible for this acceleration, is of central importance to fundamental physics. Generally, two classes of explanations have been proposed \cite{frieman2008dark}. The first invokes the presence of dark energy \cite{huterer2017dark}, an exotic component characterized by a negative pressure that effectively drives repulsive gravitational dynamics. The second attributes the observed acceleration to modifications of gravity on cosmological scales \cite{ishak2019testing}, implying a breakdown or extension of general relativity beyond its standard domain of validity. The simplest and most commonly adopted description of dark energy is provided by the cosmological constant, within the framework of general relativity, where it is interpreted as a manifestation of vacuum energy on cosmological scales. This assumption underlies the $\Lambda$ Cold Dark Matter ($\Lambda$CDM) model, which currently serves as the standard paradigm in cosmology \cite{collaboration2020planck}. Despite its phenomenological success and  simplicity, the $\Lambda$CDM model is affected by a well-known theoretical tension, usually referred to as the cosmological coincidence problem \cite{sivanandam2013cosmological, velten2014aspects}.

Within the standard $\Lambda$CDM framework, the cosmological constant represents a non-dynamical form of dark energy, with a constant energy density that does not evolve with time and space. While this minimal description successfully accounts for a wide range of cosmological observations, it leaves open fundamental theoretical questions. In particular, the assumption of a strictly constant vacuum energy gives rise to the cosmological coincidence problem, which highlights the tension between the observed late-time dominance of dark energy and the vastly different energy scales predicted by quantum field theory.

A natural extension of this framework consists in promoting dark energy to a dynamical component by introducing an additional degree of freedom, commonly realized through a scalar field $\phi$ \cite{wetterich1988cosmology, ratra1988cosmological, caldwell1998cosmological, zlatev1999quintessence}. In such scenarios, the accelerated expansion arises from the evolution of the scalar field. These models allow for a time-dependent equation of state and offer a richer phenomenology than the cosmological constant.

Recent observational advances, in particular the high-precision baryon acoustic oscillation measurements delivered by the Dark Energy Spectroscopic Instrument Data Releases DR1 and DR2 \cite{adame2025desi, DESI:2024kob, DESI:2025zgx, DESI:2025fii}, have enabled increasingly stringent constraints on the dark energy equation of state parameter $w$ \cite{dinda2024new, choudhury2024updated, chakraborty2025desi, dinda2025model, dinda2025cosmic}. When interpreted within the widely used Chevallier-Polarski-Linder (CPL) parametrization,  $w(z) = w_0 + w_a \frac{z}{1+z}$  \cite{linder2003exploring, linden2008test, scherrer2015mapping}, several analyses have reported a preference for an apparent crossing of the phantom divide  at intermediate redshifts ($z \simeq 0.5$).

Within general relativity and for a canonical scalar field, such phantom behavior is theoretically problematic, as it implies a growing dark energy density with cosmic expansion, $\dot{\rho}_{\rm de}>0$, and is typically associated with instabilities \cite{caldwell2003phantom, carroll2003can}. This tension has motivated renewed scrutiny of the physical interpretation of phenomenological parametrizations of $w(z)$.

In this context, recent work \cite{dinda2025physical} has demonstrated that physically consistent thawing quintessence models, particularly when allowing for a non-zero spatial curvature, can fit current observational data with an accuracy comparable to that of the CPL parametrization in a spatially flat Universe. These studies conclude that the inferred phantom crossing is not demanded by the data themselves, but rather reflects limitations of phenomenological parametrizations that are not anchored in an underlying physical model.

Among scalar field realizations of dynamical dark energy, quintessence has long been studied as a theoretically consistent alternative to the cosmological constant. In these models, cosmic acceleration is driven by the slow evolution of a scalar field rolling down a self-interaction potential $V(\phi)$. A wide variety of potentials has been explored in the literature, including quadratic free field potentials \cite{ratra1991expressions, urena2009dynamics}, power law potentials \cite{peebles1988cosmology, ratra1988cosmological}, exponential potentials \cite{Halliwell:1986ja, copeland1998exponential, Heisenberg:2018yae}, and hyperbolic potentials \cite{Yang:2018xah}. For example, Heisenberg et al. \cite{Heisenberg:2018yae} investigated the ability of forthcoming surveys to constrain the slope parameter $\lambda$ in an exponential potential of the form $V(\phi)\propto e^{-\lambda\phi}$, while Yang et al. \cite{Yang:2018xah} performed a comparative analysis of several potential classes using multiple observational probes. Subsequently, power law potentials have been tested against independent datasets, including luminosity measurements of HII starburst galaxies \cite{cao2020cosmological}.

Although these potentials are often treated as phenomenological choices, recent theoretical considerations suggest that they may be subject to non-trivial constraints. In particular, the swampland conjectures \cite{agrawal2018cosmological, Heisenberg:2018yae, heisenberg2019dark, andriot2019further} have been argued to impose bounds on the form of scalar field potentials, potentially disfavouring models with extremely flat potentials. While the applicability of these conjectures to late-time cosmology remains debated \cite{agrawal2019dark, raveri2019swampland}, they provide additional motivation for reconstructing the dark energy potential directly from observations rather than assuming a specific functional form.

To mitigate the biases inherent to parametric approaches, model independent reconstruction techniques provide a complementary strategy. Gaussian Processes (GP) is a machine learning approach that offers a non parametric Bayesian framework, allowing cosmological observables to be reconstructed directly from the data. This method has been widely used in cosmology to reconstruct quantities such as the Hubble parameter~\cite{shafieloo2012gaussian}, the transition redshift \cite{jesus2020gaussian}, the equation of state parameter of dark energy \cite{holsclaw2011nonparametric}, and the luminosity distance~\cite{seikel2012}, etc..

The primary objective of this work is to reconstruct the dark energy scalar field potential and its associated kinetic term directly from observational data, while explicitly accounting for the effect of spatial curvature. A number of previous studies have employed non parametric techniques to reconstruct scalar field potentials in a model independent manner~\cite{nair2014exploring, mukherjee2015reconstruction, jesus2022gaussian, elizalde2024reconstruction, 
niu2024reconstruction, gadbail2025reconstruction}. Building upon these efforts, we extend the analysis by combining the most recent measurements of the Hubble parameter $H(z)$, derived from cosmic chronometers and DESI DR2 baryon acoustic oscillations, with the Pantheon+ Type~Ia supernova compilation.

In addition to reconstructing the scalar field potential, we place particular emphasis on the kinetic term $\tau(z)$ in order to identify the redshift range over which $\tau(z)>0$. This condition is required for the physical consistency of canonical scalar field models and provides an important diagnostic of the viability of the reconstructed dynamics. To assess the impact of background cosmological assumptions, we consider two distinct choices for the matter density and spatial curvature parameters: a Planck based prior with tight constraints and a large prior that allows for greater flexibility.

To further examine the robustness of the reconstruction, we perform the analysis using two different covariance functions within the Gaussian Process framework, namely the squared exponential kernel and the Matern kernel with $\nu=9/2$. This allows us to test the sensitivity of the inferred scalar field dynamics to the assumed kernel structure. Finally, we compare the reconstructed potential and kinetic evolution with two well established theoretical benchmarks, namely power law and exponential quintessence models.

This paper is organized as follows. In Section \ref{sec2}, we present the cosmological framework and the inversion formalism used for scalar field reconstruction. Section \ref{sec3} describes the observational datasets and the reconstruction methodology. Our results are discussed in Section \ref{sec4}, and concluding remarks are given in Section \ref{sec5}.

\section{Theory}\label{sec2}\ 

In this framework, we consider a quintessence scenario in which the late-time acceleration of the Universe is driven by a minimally coupled canonical scalar field, $\phi$, evolving under a self interaction potential $V(\phi)$. The primary
objective of our reconstruction is to infer the redshift dependence of the
effective scalar field potential together with its kinetic contribution.
Specifically, we aim to reconstruct the dimensionless potential, $U(z)$, and the
dimensionless kinetic energy, $\tau(z)$, by exploiting combinations of
$H(z)$ measurements and the Pantheon+ type~Ia supernova dataset. This derivation
follows a systematic two step procedure. First, the energy density, $\rho_\phi$,
and the pressure, $P_\phi$, of the scalar field are expressed in terms of its
dynamical degrees of freedom. Second, by incorporating these effective fluid
components into the Friedmann equations, we derive the explicit relations
required to reconstruct $U(z)$ and $\tau(z)$ directly from the background
expansion history.

\subsection{Energy density and pressure} \label{subsec:scalar_field}\

In this work, we assume that the Universe is homogeneous and isotropic  on large scales. Under these symmetry assumptions, the background  spacetime is described by the Friedmann-Lema\^{i}tre-Robertson-Walker  (FLRW) metric. In cosmic time $t$, the line element takes the form \begin{equation} ds^2 = -dt^2 + a^2(t)\left[\frac{dr^2}{1-kr^2}  + r^2\left(d\theta^2 + \sin^2\theta\, d\varphi^2\right)\right], \end{equation} where $a(t)$ is the scale factor and $k = 0, \pm 1$ denotes the spatial  curvature.
The energy momentum tensor for a minimally coupled canonical scalar field
$\phi(\vec{x}, t)$ is given by \cite{ratra1988cosmological, copeland2006dynamics, niu2024reconstruction}
\begin{equation}
T_{\ \beta}^\alpha=g^{\alpha \nu} \frac{\partial \phi}{\partial x^\nu} \frac{\partial \phi}{\partial x^\beta}-g_{\ \beta}^\alpha\left[\frac{1}{2} g^{\mu \nu} \frac{\partial \phi}{\partial x^\mu} \frac{\partial \phi}{\partial x^\nu}+V(\phi)\right],
\end{equation}
where $V(\phi)$ denotes the self interaction potential associated with dark energy scalar field. Assuming that the scalar field is homogeneous at the background level, the energy momentum tensor simplifies to
\begin{equation}\label{eq2}
T_{\ \beta}^{\alpha}=g^{\alpha 0} g_{\ \beta}^0\left(\frac{\mathrm{~d} \phi}{\mathrm{d} t}\right)^2+g_{\ \beta}^\alpha\left[\frac{1}{2}\left(\frac{\mathrm{~d} \phi}{\mathrm{d} t}\right)^2-V\left(\phi\right)\right] .
\end{equation}\ 
From Eq.~(\ref{eq2}), the energy density and the pressure of the
scalar field can be directly identified. The energy density is obtained from
the time--time component of the energy momentum tensor as
$\rho_\phi = -T^{0}_{0}$. For a homogeneous background configuration, the
isotropic pressure is given by the spatial diagonal components,
$P_\phi = \frac{1}{3}T^{i}_{i}$, leading to
\begin{equation}\label{rho}
\rho_\phi  =\frac{1}{2}\dot{\phi}^2+V\left(\phi\right), 
\end{equation}

\begin{equation}\label{p}
P_\phi  =\frac{1}{2}\dot{\phi}^2-V\left(\phi\right),
\end{equation}
where the dot denotes the derivative with respect to cosmic time $t$.

\color{black}

\subsection{Dark energy scalar field potential}\ 

Within the framework of general relativity, the evolution of the Universe is governed by the Friedmann equations. At late cosmological times, the energy density of radiation is negligible compared to that of other types of densities. Therefore, neglecting the radiation component, the Friedmann equations take the form

\begin{equation}\label{Freidmann1}
H^2  =\frac{8 \pi G}{3}\left(\rho_m+\rho_\phi\right)-\frac{k}{a^2},
\end{equation}
\begin{equation}\label{Freidmann2}
\frac{\ddot{a}}{a}  =-\frac{4 \pi G}{3}\left(\rho_m+\rho_\phi+3 p_\phi\right),
\end{equation}\ 
where $\rho_m$ denotes the energy density of the pressureless matter component ($p_m = 0$).

Substituting Eqs.~(\ref{rho}) and (\ref{p}) into the Friedmann equations, Eqs.~(\ref{Freidmann1}) and (\ref{Freidmann2}) yields

\begin{equation}\label{FreidmannC1}
H^2  =\frac{8 \pi G}{3}\left[\rho_m+\frac{\dot{\phi}^2}{2}+V(\phi)\right]-\frac{k}{a^2}, 
\end{equation}
\begin{equation}\label{FreidmannC2}
\frac{\ddot{a}}{a}  =-\frac{4 \pi G}{3}\left[\rho_m+2 \dot{\phi}^2-2 V(\phi)\right].
\end{equation}
Using the relation $\frac{\ddot{a}}{a}= H^2 + \dot{H}$, the potential $V(\phi)$ can be expressed as

\begin{equation}\label{V}
V(\phi)=\frac{3 H^2+\dot{H}}{8 \pi G}-\frac{\rho_m}{2}+\frac{k}{4 \pi G a^2}.
\end{equation}

Since the observational datasets employed in the reconstruction are expressed as functions of the redshift $z$, it is convenient to rewrite the time dependence appearing in Eq.~(\ref{V}) in terms of $z$. This transformation is achieved using the relation
\begin{equation}
\frac{d}{dt} = -H(1+z)\frac{d}{dz}.
\label{eq:dt_dz}
\end{equation}
Using this relation, the scalar field potential expressed in Eq.~(\ref{V}) can be rewritten as
\begin{equation}\label{U_z}
U(z)=E^2-\frac{E(1+z)}{3} \frac{d E}{d z}-\frac{\Omega_m(1+z)^3}{2}-\frac{2 \Omega_k}{3}(1+z)^2,
\end{equation}
where $
U(\phi) \equiv 8 \pi G V(\phi)/ 3 H_0^2, E(z) \equiv H(z)/H_0$,   $\Omega_m$ and $\Omega_k$ represent the dimensionless potential, the dimensionless expansion rate, the present-day density parameters for matter and curvature, respectively.\

In contrast, by eliminating the potential term $V(\phi)$ from Eqs.~(\ref{FreidmannC1}) and~(\ref{FreidmannC2}), the kinetic energy of the scalar field, defined as $T \equiv \dot{\phi}^2/2$, can be expressed as

\begin{equation}
T=-\frac{\dot{H}}{8 \pi G}-\frac{\rho_m}{2}+\frac{k}{8 \pi G a^2},
\end{equation}
consequently, by using the dimensionless kinetic energy defined as $\tau \equiv 8\pi G T/3H_0^2 $, its evolution with respect to redshift can be written as

\begin{equation}\label{tau}
\tau(z)=\frac{E(1+z)}{3} \frac{d E}{d z}-\frac{\Omega_m(1+z)^3}{2}-\frac{\Omega_k}{3}(1+z)^2 .
\end{equation}

Eqs.~\eqref{U_z} and~\eqref{tau} allow us to reconstruct the dimensionless potential, $U(z)$, and kinetic energy, $\tau(z)$, by first reconstructing $H(z)$ and its derivative through GP using Hubble data. Alternatively, these quantities can be reconstructed using the Pantheon+ Type~Ia supernova data. For this purpose, it is convenient to express them in terms of the dimensionless transverse comoving distance\footnote{The dimensionless distances $D_i$ are related to the corresponding dimensionful distances $d_i$ through $D_i \equiv d_i/d_H$, where $d_H \equiv c/H_0$ denotes the Hubble distance.} $D_M(z)$, defined as

\begin{equation}\label{D_M}
D_M(z)=\frac{1}{\sqrt{-\Omega_k}} \sin \left(\sqrt{-\Omega_k} D_C(z)\right),
\end{equation}
where the line of sight comoving distance, $D_C(z)$, is related to the dimensionless expansion rate, $E(z)$,  through
\begin{equation}\label{dD_c}
D_C^{\prime}(z)=\frac{1}{E(z)},
\end{equation}
where the prime denotes the derivative with respect to the redshift $z$.\\

Now, taking the derivative of the transverse comoving distance $D_M(z)$, Eq.~(\ref{D_M}), with respect to redshift $z$, we obtain

\begin{equation}\label{dD_M}
D_M^{\prime}(z)=\cos \left(\sqrt{-\Omega_k} D_C(z)\right) D_C^{\prime}(z).
\end{equation}\
Combining Eqs. (\ref{D_M}) and (\ref{dD_M}), we find 
\begin{equation}\label{DMC}
\left(\frac{D_M^{\prime}}{D_C^{\prime}}\right)^2-\Omega_k D_M^2=1.
\end{equation}
This equation can then be rewritten, using Eq.~(\ref{dD_c}), in terms of the dimensionless expansion rate $E(z)$ as

\begin{equation}\label{EE}
E^2=\frac{1+\Omega_k D_M^2}{D_M^{\prime 2}}.
\end{equation}\
Differentiating Eq.~(\ref{EE}) with respect to redshift $z$, we obtain

\begin{equation}
E \frac{d E}{d z}=\frac{\Omega_k D_M\left(D_M^{\prime 2}-D_M D_M^{\prime \prime}\right)-D_M^{\prime \prime}}{D_M^{\prime 3}},
\end{equation}
which enables the dimensionless potential, defined in Eq.~(\ref{U_z}) to be expressed in terms of the transverse comoving distance, $D_M(z)$, and its first and second derivatives, $D_M'(z)$ and $D_M''(z)$, respectively.

\begin{equation}\label{U_D}
U(z)=\frac{1+\Omega_k D_M^2}{D_M^{\prime 2}}+\left(\frac{1+z}{3}\right) \frac{D_M^{\prime \prime}+\Omega_k D_M\left(D_M D_M^{\prime \prime}-D_M^{\prime 2}\right)}{D_M^{\prime 3}}-\frac{\Omega_m(1+z)^3}{2}-\frac{2 \Omega_k}{3}(1+z)^2 .
\end{equation}\
Similarly, the dimensionless kinetic energy, can be expressed in terms of the transverse comoving distance as

\begin{equation}\label{tau_D}
\tau(z)=\left(\frac{1+z}{3}\right) \frac{\Omega_k D_M\left(D_M^{\prime 2}-D_M D_M^{\prime \prime}\right)-D_M^{\prime \prime}}{D_M^{\prime 3}}-\frac{\Omega_m(1+z)^3}{2}-\frac{\Omega_k}{3}(1+z)^2 .
\end{equation}

\section{Dataset and Methodology}\label{sec3}

\subsection{Dataset}\label{sec3_1}\

The observational data sets used in this work are divided into two independent categories: 
(i) recent measurements of the Hubble parameter, $H(z)$, and 
(ii) Type~Ia supernova observations drawn from the Pantheon$+$ compilation.

The $H(z)$ measurements considered here are obtained using two complementary techniques. 
The first is the differential age method, commonly referred to as the cosmic chronometer (CC) approach, which relies on the relation between the Hubble parameter and the time derivative of the redshift of passively evolving galaxies, expressed as~\cite{Jimenez2002}
\begin{equation}
H(z) = -\frac{1}{1+z}\frac{dz}{dt}.
\label{eq:CC}
\end{equation}
This relation allows the estimation of $H(z)$ through relative age differences measured in massive early type galaxies at different redshifts. 
The second technique exploits the clustering of galaxies and quasars, where the position of the baryon acoustic oscillation (BAO) peak along the line of sight direction provides a direct probe of the expansion rate~\cite{Gaztanaga2009}.

For our analysis, we utilize 32 measurements from the CC method together with six recent points from the Dark Energy Spectroscopic Instrument (DESI) DR2  \cite{DESIDR2}. 
The DESI DR2 measurements are incorporated exclusively in the form of the ratio $D_H/r_d$, obtained from luminous red galaxies (LRG), emission line galaxies (ELG), quasars (QSO), and Lyman-$\alpha$ forest (Ly$\alpha$) samples at effective redshifts 
$z_{\mathrm{eff}} = (0.510, 0.706, 0.934, 1.321, 1.484, 2.330)$. To convert these measurements into estimates of the Hubble parameter $H(z)$, an assumption about the sound horizon scale, $r_d$, is required. This follows from the relation $D_H = c/H(z)$, implying that the inferred $H(z)$ values depend on the choice of $r_d$. In this work, we consider values of  $r_d$ from the Planck 2018 analysis, i.e. $r_d = 147.09 \pm 0.26$~Mpc, $147.21 \pm 0.23$~Mpc, and $147.05 \pm 0.30$~Mpc~\cite{collaboration2020planck}, as well as two alternative determinations obtained from a sound-horizon agnostic analysis combining different observational datasets~\cite{GarciaEscudero2026}, i.e. $r_d = 147.70 \pm 0.17$~Mpc and $144.80 \pm 1.60$~Mpc. For each case, we evaluate the likelihood and identify which value of $r_d$ provides the best fit to the data.

Since CC measurements are not statistically independent, the corresponding covariance matrices are properly taken into account in our analysis to ensure the robustness of the reconstruction. The dominant sources of uncertainty arise from factors such as stellar metallicity and contamination by younger stellar populations, which are typically uncorrelated across galaxies at different redshifts. Additional systematic uncertainties originate from the adoption of a common stellar population synthesis model, particularly through assumptions related to the initial mass function and stellar libraries. 
A detailed discussion of these effects can be found in Ref.~\cite{Moresco2020}. A summary of the full $H(z)$ data set is provided in Table~\ref{table1}.

We also include the Pantheon$+$ Type~Ia supernova compilation~\cite{Brout:2021mpj}, which consists of 1701 measurements obtained from 1550 supernovae covering the redshift range $0.00122 \leq z \leq 2.26137$. 
To mitigate potential biases in the inference of the background expansion arising from local peculiar velocities, we restrict the Pantheon$+$ sample to redshifts at which the global Hubble flow is expected to dominate. 
Specifically, we impose a lower redshift cutoff at $z = 0.01$, since stochastic motions within the local volume can significantly distort the distance-redshift relation below this threshold. 
This selection removes 111 events, yielding a final cosmological sample of 1590 supernova light curves spanning the interval $0.01016 \leq z \leq 2.26137$. The corresponding data products, including the full covariance matrix, are publicly available through the Pantheon$+$ GitHub repository\footnote{See \url{https://github.com/PantheonPlusSH0ES/DataRelease}}.

\begin{table}[!t]
	\centering
	\caption{Compilation of $H(z)$ measurements with their corresponding uncertainties $\sigma_H$, expressed in km s$^{-1}$ Mpc$^{-1}$, used in the present study.}
	\label{table1}
	\begin{tabular}{|c|c|c|c|c|c|c|c|}
		\hline
		\multicolumn{8}{|c|}{CC data} \\
		\hline
		\textbf{$z$} & \textbf{$H(z)$} & \textbf{$\sigma_{H}$} & Ref. & \textbf{$z$} & \textbf{$H(z)$} & \textbf{$\sigma_{H}$} & Ref. \\
		\hline
		0.07  & 69.0  & 19.6  & \cite{zhang2014four1} &  0.4783 & 80.9  & 9 & \cite{Moresco2016mzx}\\
		0.09  & 69.0  & 12.0  & \cite{jimenez2003constraints}  & 0.48  & 97.0  & 62.0  & \cite{stern2010cosmic1} \\
		0.12  & 68.6  & 26.2  & \cite{zhang2014four1} &  0.5929 & 104 & 13 & \cite{moresco2012improved} \\
		0.17  & 83.0  & 8.0   & \cite{stern2010cosmic}   & 0.6797 & 92.0  & 8.0 &  \cite{moresco2012improved} \\
		0.1791 & 75.0  & 4.0  & \cite{moresco2012improved}& 0.75 &  98.8 & 33.6 & \cite{borghi2022toward}\\
		0.1993 & 75.0  & 5.0  & \cite{moresco2012improved} & 0.7812 & 105  & 12.0 &  \cite{moresco2012improved}\\
		0.2  & 72.9  & 29.6  & \cite{zhang2014four1}  & 0.8754 & 125.0 & 17.0 &  \cite{moresco2012improved} \\
		0.27  & 77.0  & 14.0  & \cite{stern2010cosmic}  & 0.88  & 90.0  & 40.0  & \cite{stern2010cosmic1}   \\
		0.28  & 88.8  & 36.6  & \cite{zhang2014four1}  & 0.90  & 117.0 & 23.0  & \cite{stern2010cosmic}  \\
		0.3519 & 83.0  & 14.0 & \cite{moresco2012improved} &  1.037  & 133.5 & 17.6 &\cite{moresco2012improved}\\
		0.3802 & 83.0  & 13.5 & \cite{Moresco2016mzx} & 1.30  & 168.0 & 17.0  & \cite{stern2010cosmic} \\
		0.4  & 95.0  & 17.0  & \cite{stern2010cosmic}  & 1.363  & 160.0 & 33.8 & \cite{Moresco2015cya} \\
		0.4004 & 77.0  & 10.2 & \cite{Moresco2016mzx} & 1.43  & 177.0 & 18.0  & \cite{stern2010cosmic}  \\
		0.4247 & 87.1  & 11.2 & \cite{Moresco2016mzx} & 1.53  & 140.0 & 14.0  & \cite{stern2010cosmic}\\
		0.4497 & 92.8  & 12.9 & \cite{Moresco2016mzx} & 1.75  & 202.0 & 40.0  & \cite{stern2010cosmic}\\
		0.47  & 89.0  & 49.6  & \cite{ratsimbazafy2017age} & 1.965  & 186.5 & 50.4 & \cite{Moresco2015cya}  \\
		\hline
		\multicolumn{8}{|c|}{DESI DR2 data} \\
		\hline
		\textbf{$z$} & \textbf{$H(z)$} & \textbf{$\sigma_{H}$} & Ref. & \textbf{$z$} & \textbf{$H(z)$} & \textbf{$\sigma_{H}$} & Ref. \\
		\hline
		0.51 &  94.70 & 2.12  & \cite{DESIDR2}& 1.321 & 146.05  &  2.79  & \cite{DESIDR2}  \\
		0.706 & 106.42 & 2.15  & \cite{DESIDR2} & 1.484 & 161.53  &  6.74 & \cite{DESIDR2}  \\
		0.934 & 117.36  &  1.82  & \cite{DESIDR2} &2.330 &  239.85  &  3.86 & \cite{DESIDR2} \\
		\hline
	\end{tabular}
\end{table}

\subsection{Methodology}\ 

GP regression provides a flexible Bayesian framework for reconstructing functions and their derivatives in a model independent manner from observational data points~\cite{seikel2012}. 
Formally, GP is defined as a collection of random variables for which any finite subset follows a joint multivariate Gaussian distribution~\cite{williams2006gaussian}. A key advantage of GP over other non parametric methods is that it provides both the posterior mean function and the associated uncertainty at each redshift, once a covariance function (or kernel) is specified.

The statistical properties of the reconstructed function, $f(z)$, are completely specified by a mean function, $\mu(z)$, and a covariance function $k(z,z')$. The kernel determines the correlation structure between function values evaluated at different redshifts and controls the degree of smoothness of the reconstruction, ensuring that the inferred trajectory represents a correlated stochastic function rather than a set of independent data points.

Within the GP framework, the observational data are interpreted as noisy realizations of the underlying latent function, $f(z)$, whose posterior distribution is characterized by a redshift dependent mean and variance. 
Correlations between reconstructed values at two redshifts, $z$ and $z^{\prime}$, are entirely governed by the choice of covariance function $k(z,z^{\prime})$. 
A variety of kernel families have been explored in the literature, including squared exponential, Matern, rational quadratic, and periodic kernels \cite{williams2006gaussian, seikel2012}. 
Among these, a commonly adopted choice is the squared exponential kernel, also known as the radial basis function (RBF), given by
\begin{equation}\label{RBF}
k\left(z, z^{\prime}\right)=\sigma_f^2 \exp \left[-\frac{\left(z-z^{\prime}\right)^2}{2 \ell^2}\right]
\end{equation}
where $\sigma_f$ controls the overall amplitude of the correlated signal and $\ell$ sets the characteristic correlation length, governing the smoothness of the reconstructed function. 
The optimal values of these hyperparameters are determined by maximizing the marginal likelihood, thereby allowing the data to directly inform the functional form of the reconstruction.

In addition to the squared exponential kernel, we also consider the Matern covariance function with $\nu = 9/2$, which has been widely employed in cosmological reconstruction analyses. 
This kernel yields sample functions that are four times mean-square differentiable, ensuring that second and, in principle, higher-order derivatives of the inferred function can be reconstructed reliably.
The Matern $(\nu=9/2)$ kernel is given by
\begin{equation}\label{matern}
\begin{aligned}
k(z,z^\prime) = \sigma_f^2 \exp\!\left(-\frac{3|z-z^{\prime}|}{\ell}\right)
\Bigg[ & 1 + \frac{3|z-z^{\prime}|}{\ell}
+ \frac{27(z-z^{\prime})^2}{7\ell^2} \\
& + \frac{18|z-z^{\prime}|^3}{7\ell^3}
+ \frac{27(z-z^{\prime})^4}{35\ell^4} \Bigg],
\end{aligned}
\end{equation}
which helps improve numerical stability in reconstructing higher-order derivatives compared to kernels with lower differentiability (smaller $\nu$).\ 

In this work, we aim to reconstruct the dimensionless scalar field potential, and the associated kinetic term, directly from cosmological observations in a model independent manner. To this end, we employ GP techniques, which allow the redshift dependence of the relevant cosmological quantities to be inferred from the data without assuming a specific parametric form for the scalar field potential.

For the datasets described above, we make use of the publicly available \text{GaPP} (Gaussian Processes in Python) package~\cite{seikel2012}. The reconstruction is performed following two complementary approaches based on different cosmological observables.

In the first approach, Eqs.~(\ref{U_z}) and~(\ref{tau}) express the dimensionless potential and the kinetic term in terms of the dimensionless expansion rate, $E(z)$, and its first derivative with respect to redshift. Consequently, GP reconstruction of $H(z)$  and its derivative enables inference of both quantities. This reconstruction is carried out using the Hubble parameter measurements described in Sec.~\ref{sec3_1}.

Alternatively, Eqs.~(\ref{U_D}) and~(\ref{tau_D}) show that the same physical quantities can be written in terms of the dimensionless transverse comoving distance, $D_M(z)$, and its derivatives. Using the Pantheon+ Type~Ia supernova apparent magnitude measurements, $m_B$, we reconstruct $D_M(z)$ and its derivatives through GP regression. This second approach therefore allows for an independent determination of the scalar field potential and kinetic term based solely on supernova data.

The reconstruction of the dimensionless scalar field potential, and the associated kinetic term, depends explicitly on the assumed values of the matter density parameter, $\Omega_m$, and the spatial curvature, $\Omega_k$, as can be seen from Eqs.~(\ref{U_z}), (\ref{tau}), (\ref{U_D}), and (\ref{tau_D}). Since GP reconstructions alone do not provide intrinsic constraints on these background cosmological parameters, their values must be specified through external priors inferred from independent observations. To account for this dependence and to test the robustness of our results, we follow the strategy proposed by Jesus et al.~\cite{jesus2022gaussian} and consider two distinct prior choices.

The first prior set is based on the Planck 2018 results~\cite{collaboration2020planck}, adopting  $\Omega_m = 0.315 \pm 0.022$ and $\Omega_k = -0.011 \pm 0.019$, both taken at the $3\sigma$ confidence level. The use of a $3\sigma$ level is motivated by the fact that these constraints were obtained assuming the $\Lambda$CDM model, rather than a dynamical scalar field dark energy cosmologies. The second prior set corresponds to a large prior, defined by $\Omega_m = 0.30 \pm 0.05$  and $\Omega_k = 0.0 \pm 0.05$. This choice is sufficiently conservative to encompass the range of values allowed by current cosmological observations and allows us to assess the sensitivity of the reconstruction to the assumed background parameters.

In order to assess the robustness of the reconstruction, we perform the analysis using two different covariance functions, defined in Eqs. (\ref{RBF}) and (\ref{matern}), within the GP framework to examine whether the choice of kernel significantly affects the reconstruction of  $U(z)$ and $\tau(z)$.

In practice, this reconstruction is carried out through a parallelized numerical pipeline, which samples the multivariate Gaussian distributions associated with the GP reconstructions of $H(z)$ and its derivative, as well as of $D_M(z)$ and its derivatives. This procedure enables the consistent inference of the dimensionless scalar field potential,  and the corresponding kinetic term.

As a consistency check, we compare the reconstructed scalar field potential, with two well established quintessence models commonly used as theoretical benchmarks: the Ratra-Peebles power law potential \cite{peebles1988cosmology, ratra1988cosmological}, and the exponential potential \cite{Halliwell:1986ja, copeland1998exponential}.

The Power Law potential is given by
\begin{equation}
    U_{\rm PL}(\Phi) = \frac{\kappa}{2}\,\Phi^{-\alpha},
\end{equation}
where $\Phi$ denotes the dimensionless scalar field, related to the physical field $\phi$ through $\Phi \equiv \sqrt{\frac{8\pi G}{3}}\,\phi$, and $\kappa$ can be expressed as a function of $\alpha$ as 
\begin{equation}
    \kappa = \frac{8}{3}\left(\frac{\alpha+4}{\alpha+2}\right)
    \left[\frac{2}{3}\,\alpha(\alpha+2)\right]^{\alpha/2},
\end{equation}
where $\alpha$ is a free parameter. We also consider the Exponential potential, another widely studied realization of quintessence, defined as
\begin{equation}
    U_{\rm EXP}(\Phi) = U_0\,e^{-\lambda\Phi},
\end{equation}
where $\lambda$ is a free parameter that characterizes the slope of the potential in field space.

In the limiting cases $\alpha \to 0$ and $\lambda \to 0$, both potentials approach a constant value, $U(\Phi)\equiv\text{const}$. The corresponding cosmological dynamics then reproduce those of the standard $\Lambda$CDM model.

\color{black}

\begin{figure}[!t]
\begin{minipage}{0.5\textwidth}
\includegraphics[width=\linewidth]{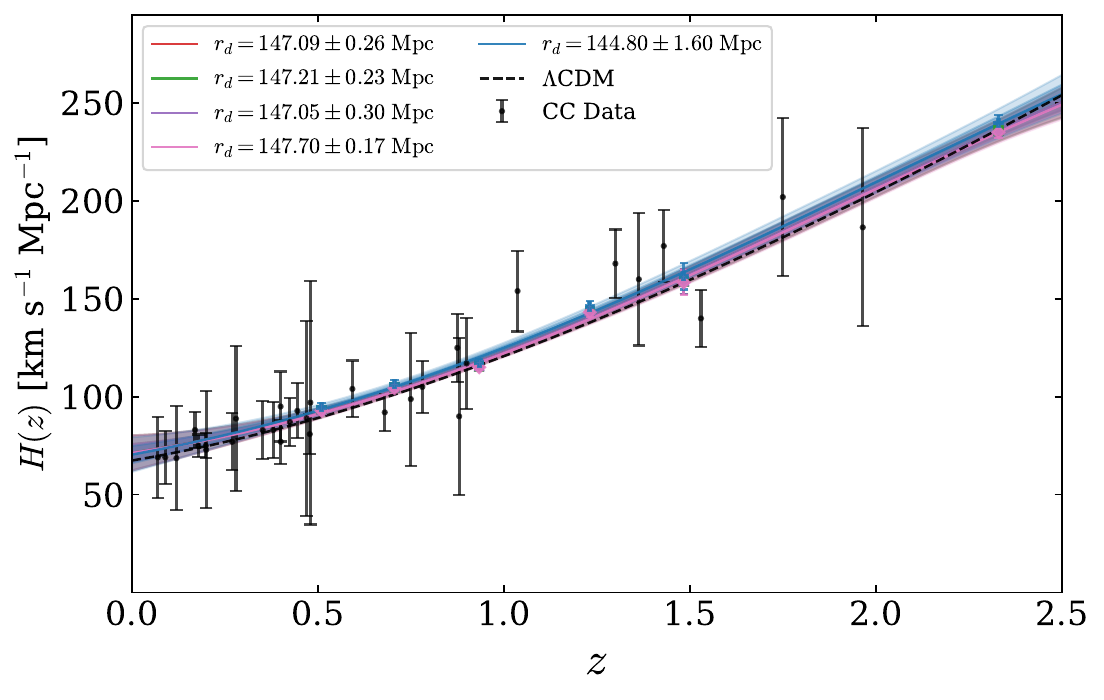}
\end{minipage}%
\begin{minipage}{0.5\textwidth}
\includegraphics[width=\linewidth]{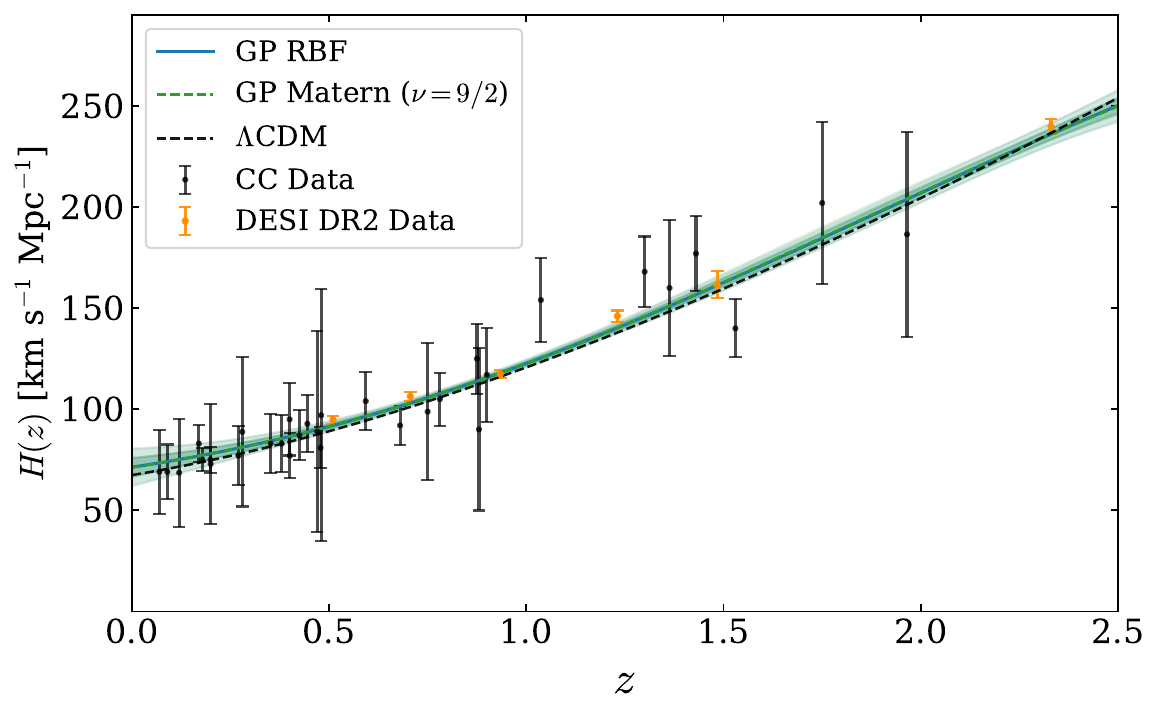}
\end{minipage}
\caption{Reconstruction of the Hubble parameter $H(z)$ from the combined CC + DESI DR2 dataset. 
The left panel shows the reconstructions obtained with the squared exponential (RBF) kernel for five different values of $r_d$. The right panel presents reconstructions obtained using both the RBF and Matern ($\nu = 9/2$) kernels, adopting the fiducial value $r_d = 144.8 \pm 0.17$~Mpc. 
In both panels, the black dashed curve corresponds to the $\Lambda$CDM model, and the dark and light shaded regions denote the $1\sigma$ (68\%) and $2\sigma$ (95\%) confidence intervals, respectively.}
\label{Fig1}  

\vspace{1em} 

\begin{minipage}{0.5\textwidth}
\includegraphics[width=\linewidth]{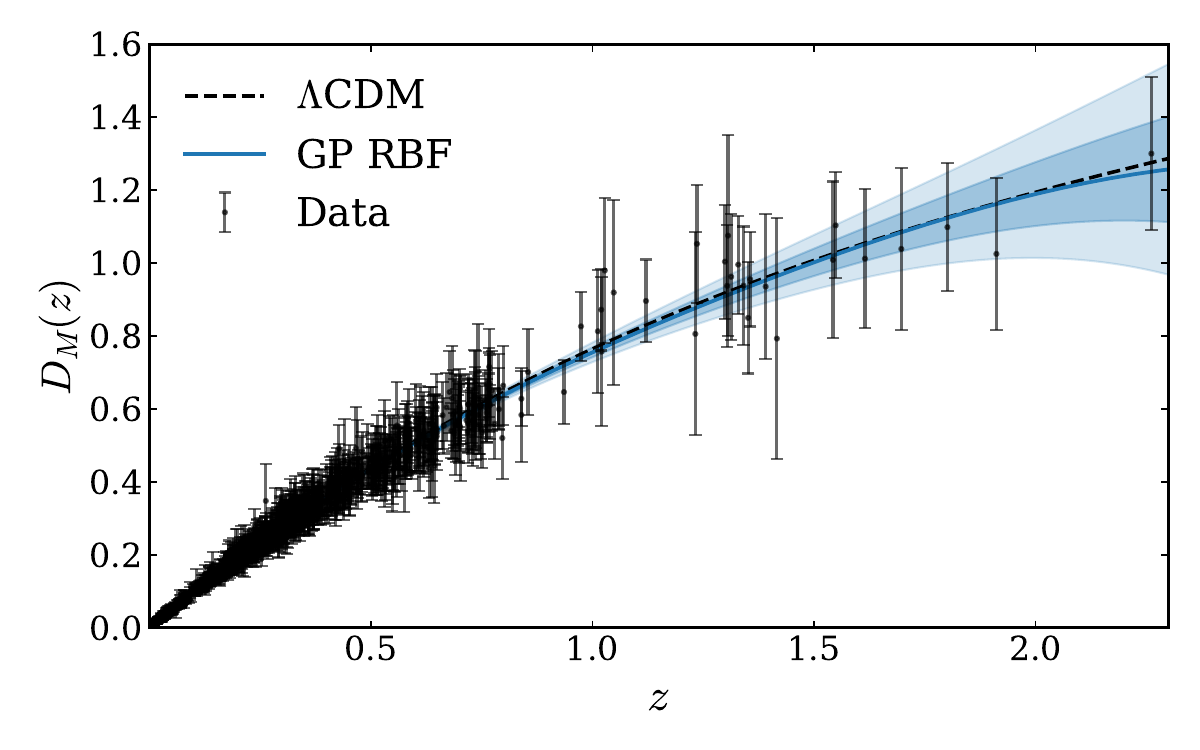}
\end{minipage}%
\begin{minipage}{0.5\textwidth}
\includegraphics[width=\linewidth]{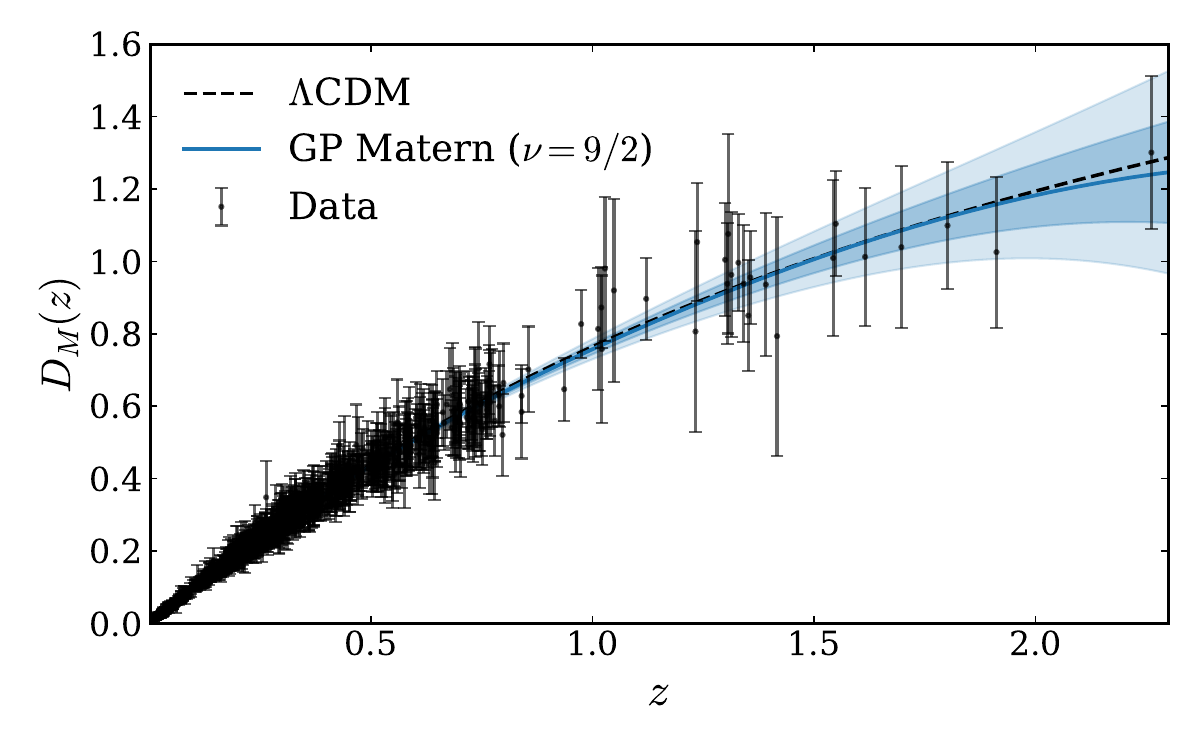}
\end{minipage}

\caption{Reconstruction of the dimensionless transverse comoving distance $D_M(z)$ from Type~Ia supernova data. The left panel uses the squared exponential (RBF) kernel, while the right panel uses the Matern ($\nu=9/2$) kernel. In both panels, the black dashed curve represents the $\Lambda$CDM model. The blue and light blue shaded regions denote the 1$\sigma$ (68\%) and 2$\sigma$ (95\%) confidence intervals, respectively.}

\label{Fig2}  
\end{figure}

\section{Results and Discussions}\label{sec4} \

In this section, we present and discuss our results, which are organized into two primary cases based on the choice of prior: (i) the Planck prior and (ii) a Large prior for the cosmological density parameters. As noted previously, the dimensionless scalar field potential, $U(z)$, and the kinetic energy, $\tau(z)$, defined in Eqs.~(\ref{U_z}) and (\ref{tau}), depend on the expansion rate, $E(z)$, and its first derivative. Likewise, according to Eqs.~(\ref{U_D}) and (\ref{tau_D}), these quantities depend on the dimensionless transverse comoving, $D_M(z)$, and its derivatives. Therefore, in order to derive $U(z)$ and $\tau(z)$, we must first reconstruct $H(z)$ and $D_M(z)$, along with their derivatives, from the observational datasets.

Figure \ref{Fig1} shows the reconstruction of the Hubble parameter, $H(z)$, using cosmic chronometers measurements together with DESI DR2 BAO data. The left panel presents the reconstruction obtained for different values of the sound horizon, $r_d$, using the squared exponential kernel. For both the squared exponential and Matern ($\nu = 9/2$) kernels, we find that the value of $r_d$ that maximizes the marginal log-likelihood\footnote{The marginal log-likelihood reads
\begin{equation*}
\ln p(\mathbf{y}\mid\boldsymbol{\theta}) = 
-\frac{1}{2}(\mathbf{y}-\boldsymbol{\mu})^{\top}
\bigl[K_{XX} + C\bigr]^{-1}(\mathbf{y}-\boldsymbol{\mu})
-\frac{1}{2}\ln\bigl|K_{XX}+C\bigr|
-\frac{n}{2}\ln(2\pi),
\end{equation*}
where $y$ denotes the observed data, $\mu$ the prior mean, $K_{XX}$ the kernel covariance matrix evaluated at redshifts $X=\{z_i\}$, $C$ the covariance matrix of the
data, and $n$ the number of data points. This criterion balances data fit and reconstruction regularization.} is $r_d = 144.80 \pm 1.6\,\mathrm{Mpc}$, the corresponding log-likelihood values are listed in Table \ref{tab2}. The right panel displays the reconstruction obtained for this best-fit value of $r_d$ using the both kernels. Fig.  \ref{Fig2} shows the reconstructed dimensionless transverse comoving distance, $D_M(z)$, from the Pantheon+ dataset using the same kernels. In both cases, the reconstructed quantities $H(z)$ and $D_M(z)$ are in good agreement with the standard $\Lambda$CDM model.

\begin{figure}[t]
\begin{minipage}{0.51\textwidth}
\includegraphics[width=\linewidth]{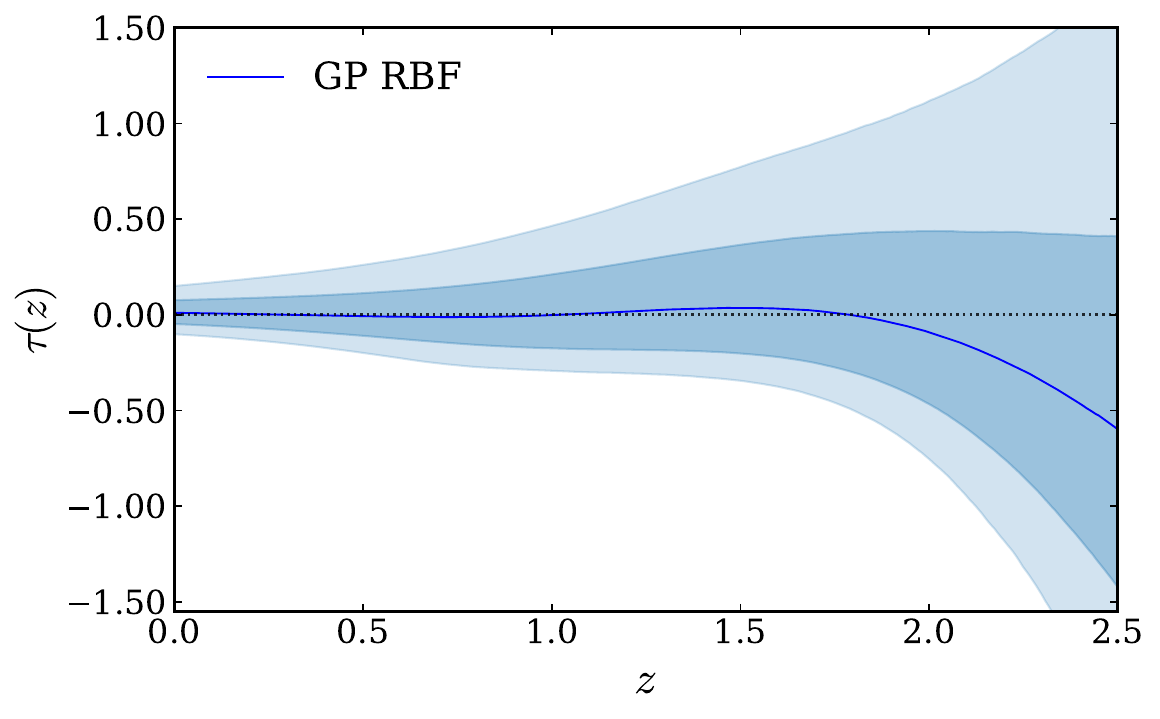}
\end{minipage}%
\begin{minipage}{0.51\textwidth}
\includegraphics[width=\linewidth]{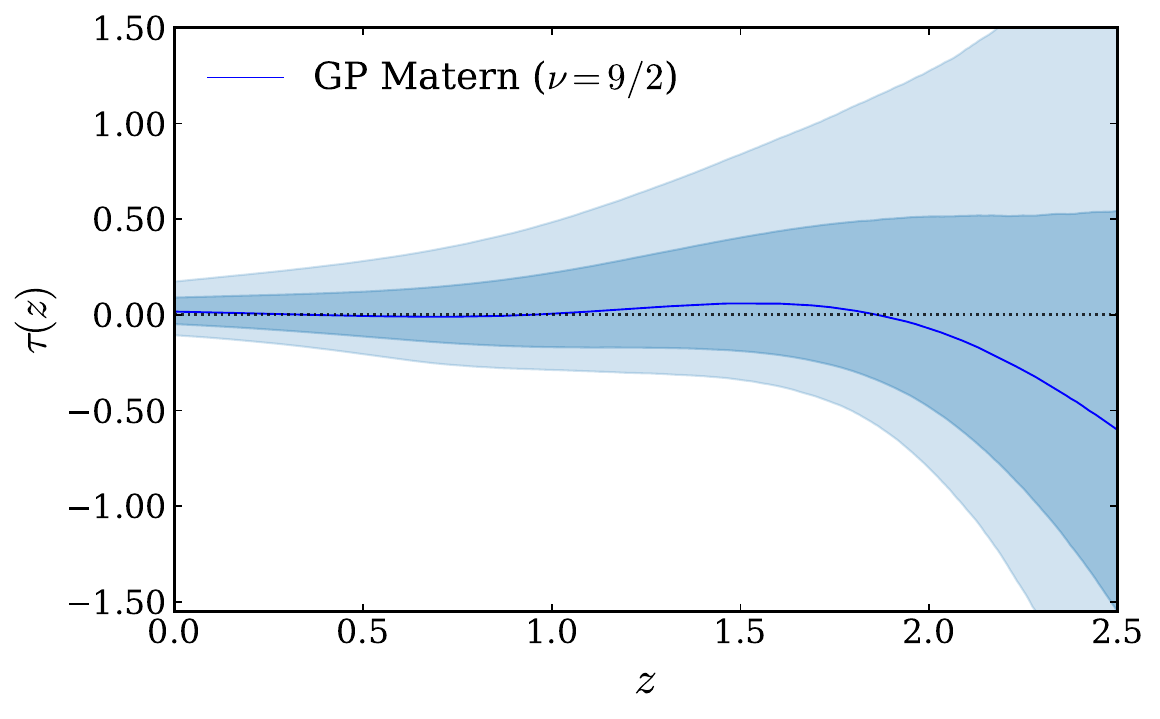}
\end{minipage}

\begin{minipage}{0.51\textwidth}
\includegraphics[width=\linewidth]{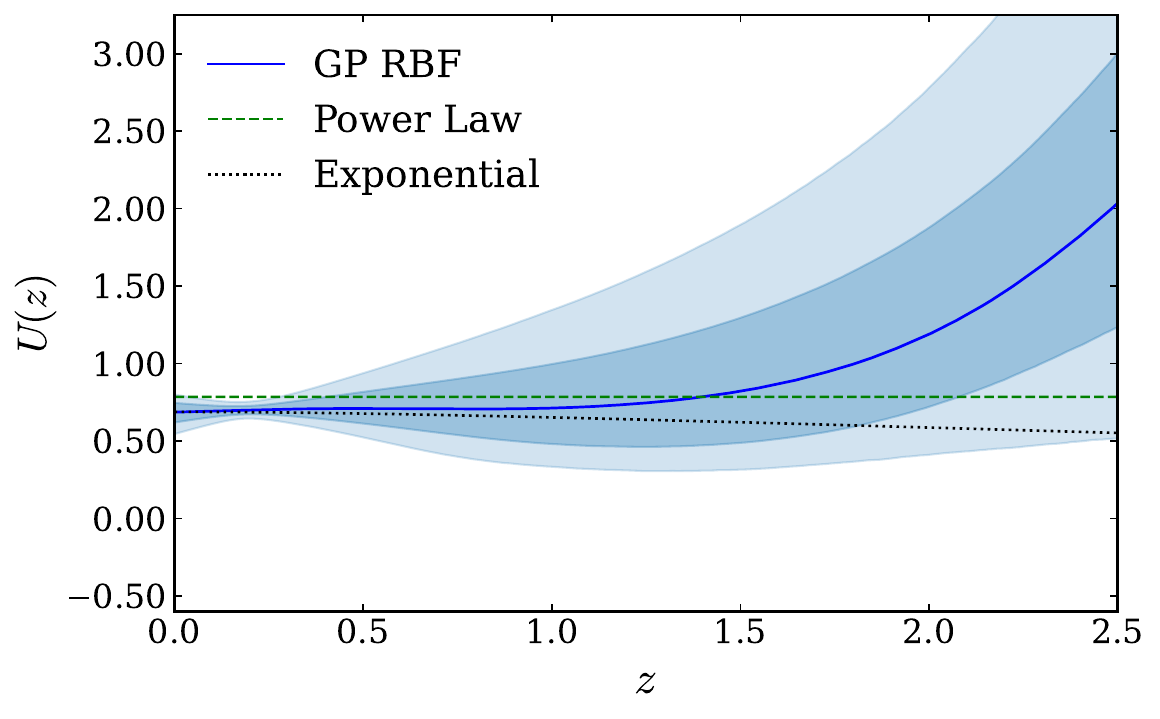}
\end{minipage}%
\begin{minipage}{0.51\textwidth}
\includegraphics[width=\linewidth]{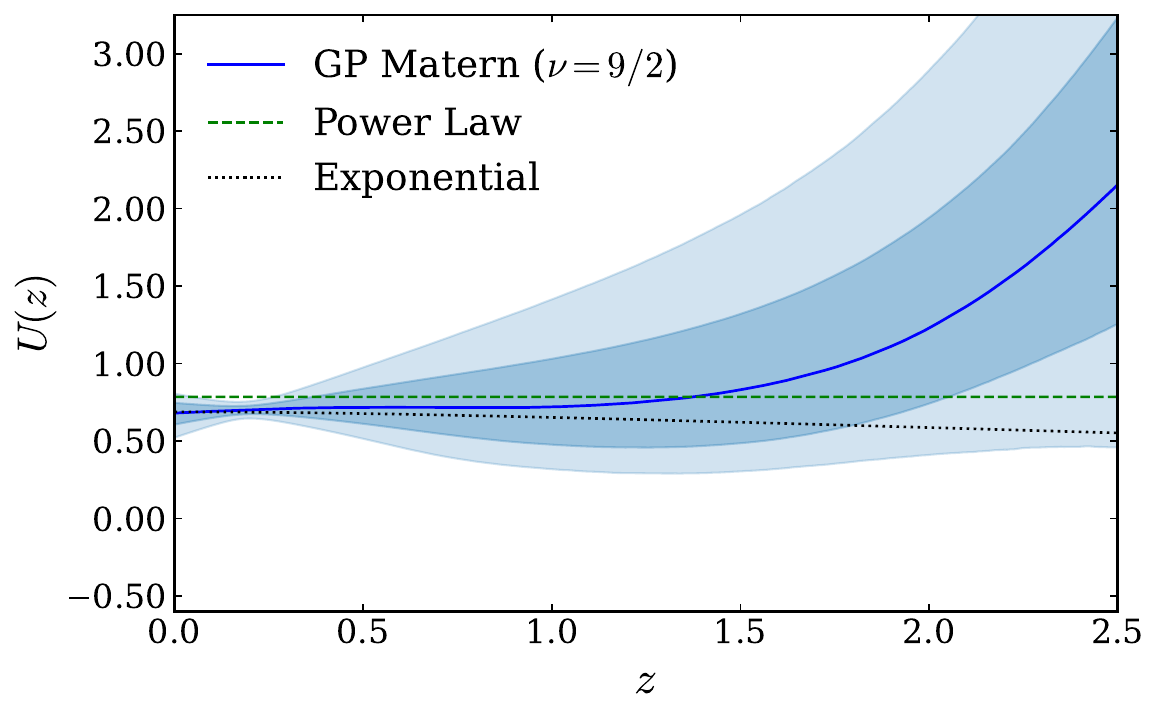}
\end{minipage}

\caption{Reconstruction of the dimensionless kinetic energy $\tau(z)$ (top panels) and the scalar field potential $U(z)$ (bottom panels) from Hubble parameter measurements data, assuming a Planck $3\sigma$ prior on $(\Omega_m,\Omega_k)$. The left panels correspond to the squared exponential (RBF) kernel, while the right panels correspond to the Matern ($\nu=9/2$) kernel. The blue and light blue shaded regions denote the 1$\sigma$ and 2$\sigma$ confidence intervals, respectively. The green dashed and black dotted curves represent the power law and exponential potentials, respectively.}

\label{Fig3}
\end{figure}\

\begin{figure}[!t]
\begin{minipage}[t]{0.53\textwidth}
\includegraphics[width=\linewidth]{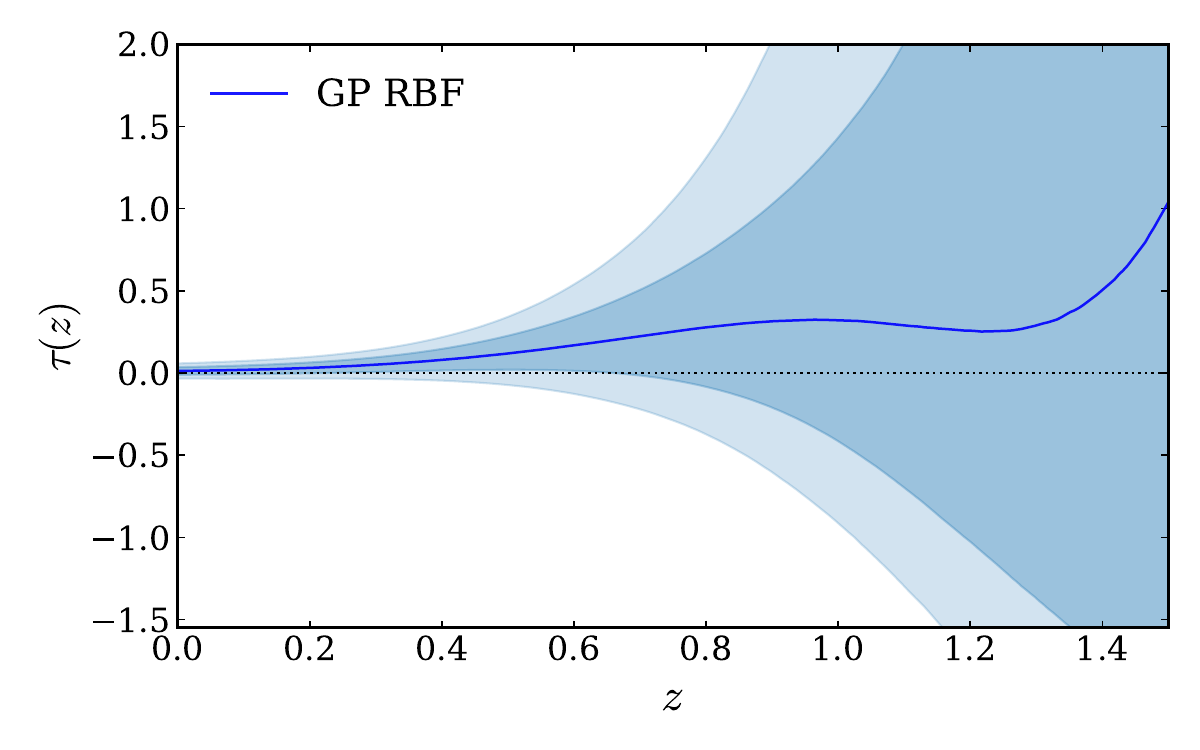}
\end{minipage}%
\hspace{-0. cm}
\begin{minipage}[t]{0.53\textwidth}
{\includegraphics[width=\linewidth]{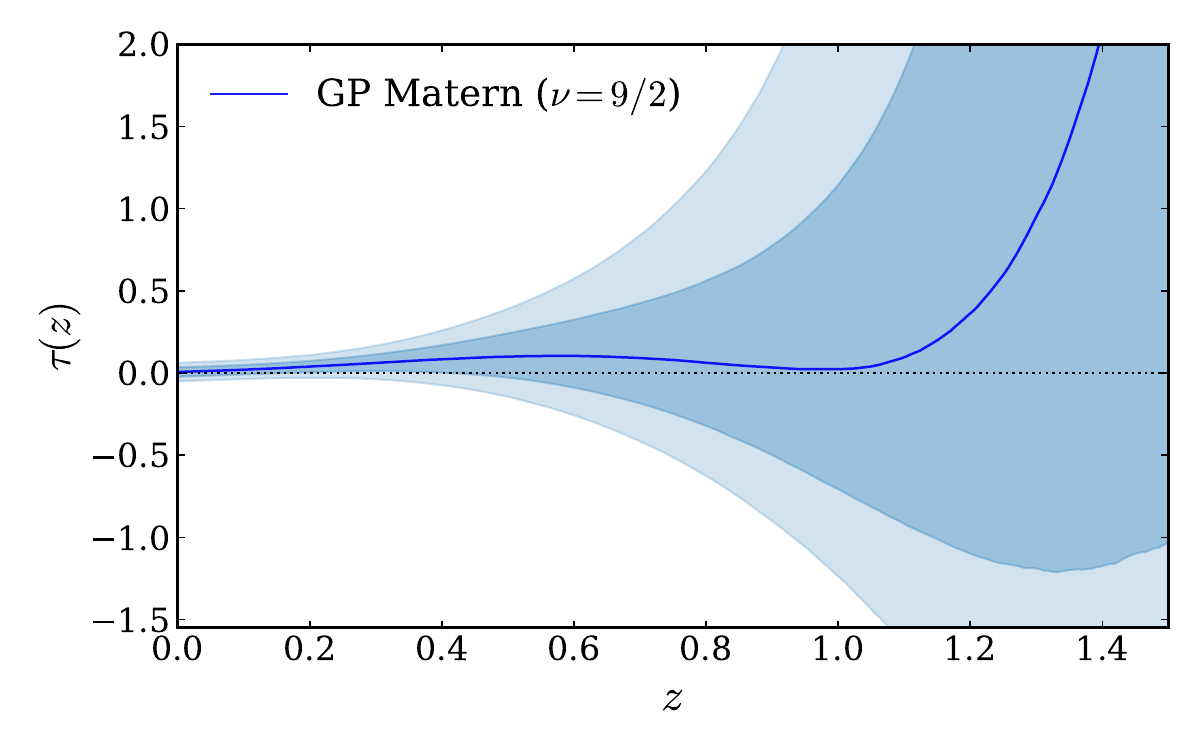}}
\end{minipage}

\begin{minipage}[t]{0.56\textwidth}
\includegraphics[width=\linewidth]{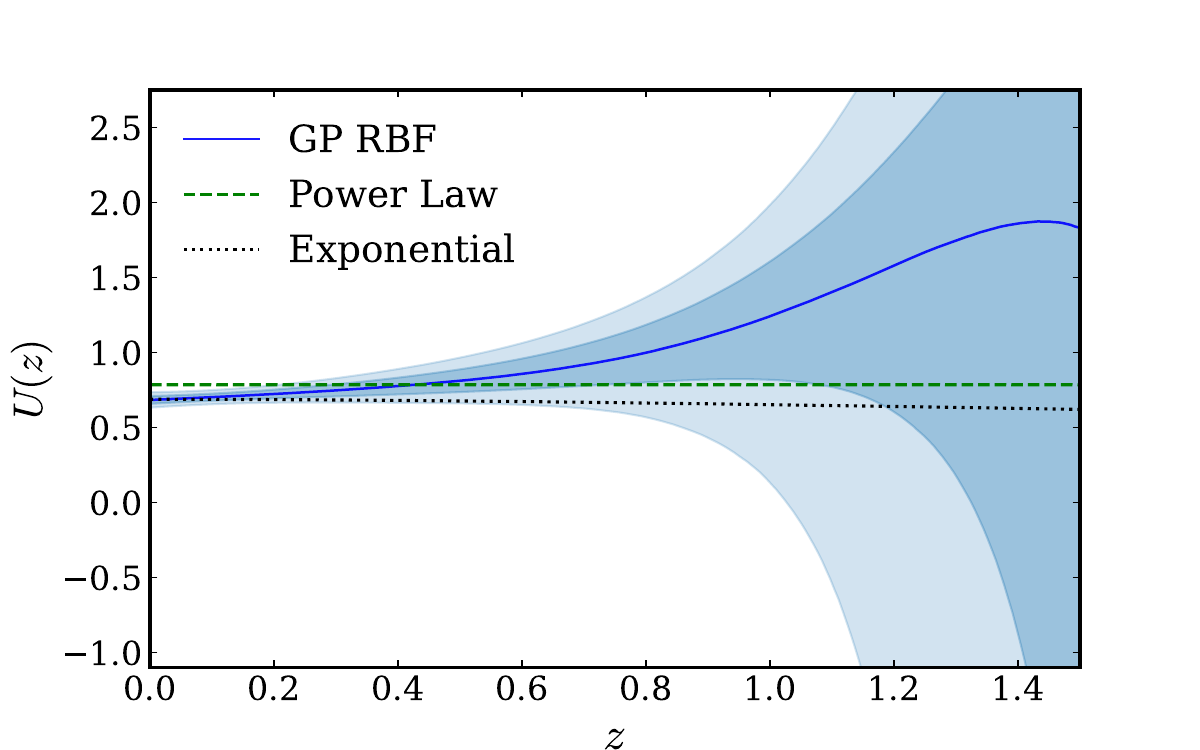}
\end{minipage}%
\hspace{-0.7 cm}
\begin{minipage}[t]{0.56\textwidth}
\includegraphics[width=\linewidth]{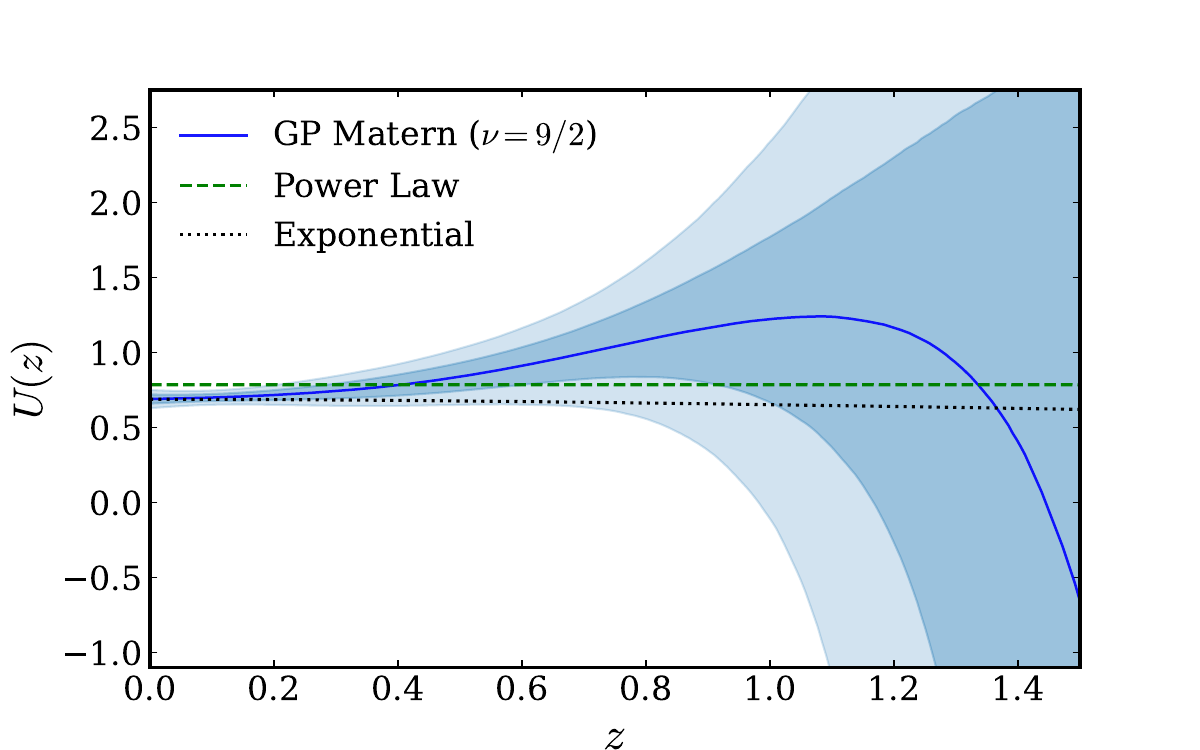}
\end{minipage}

\caption{Reconstruction of the dimensionless kinetic energy $\tau(z)$ (top panels) and the scalar field potential $U(z)$ (bottom panels) from Pantheon$+$ data, assuming a Planck $3\sigma$ prior on $(\Omega_m,\Omega_k)$. The left panels correspond to the squared exponential (RBF) kernel, while the right panels correspond to the Matern ($\nu=9/2$) kernel. The blue and light blue shaded regions denote the 1$\sigma$ and 2$\sigma$ confidence intervals, respectively. The green dashed and black dotted curves represent the power law and exponential potentials, respectively.}

\label{Fig4}
\end{figure}

\begin{table}[ht]
\centering
\caption{Marginal log-likelihoods $\ln \mathcal{L}_{\mathrm{RBF}}$ and $\ln \mathcal{L}_{\mathrm{Matern}}$ for different values of $r_d$.}
\label{tab2}
\begin{tabular}{lcccc}
\hline\hline
 $r_d$ [Mpc] & $\ln\mathcal{L}_{\mathrm{RBF}}$ & $\ln\mathcal{L}_{\mathrm{Matern}}$ \\
\hline
 147.09 $\pm$ 0.26 & -157.167 & -157.145 \\
 147.21 $\pm$ 0.23 & -157.168 & -157.145 \\
 147.05 $\pm$ 0.30 & -157.154 & -157.132 \\
 147.70 $\pm$ 0.17 & -157.147 & -157.121 \\
 144.80 $\pm$ 1.60 & -156.382 & -156.392 \\
\hline\hline
\end{tabular}
\end{table}

\subsection*{A. Reconstruction of $U(z)$ and $\tau(z)$ Using the Planck Prior}
\phantomsection
\label{secA} \

In this subsection, we present the reconstructions of $U(z)$ and $\tau(z)$ obtained using the Planck prior. First, by substituting the reconstructed $H(z)$ and its derivative in Eqs.~(\ref{U_z}) and (\ref{tau}), we perform a model independent reconstruction of $U(z)$ and $\tau(z)$ from the Hubble data. These results are shown in Fig. \ref{Fig3}, the left panels correspond to the squared exponential kernel, while the right panels correspond to the Matern ($\nu = 9/2$) kernel.

We find that the median reconstructed kinetic term, $\tau(z)$, crosses zero and becomes negative in the intervals $0.28 < z < 1.02$ and $z > 1.78$ for the squared-exponential kernel, and in the intervals $0.34 < z < 0.95$ and $z > 1.86$ for the Matern ($\nu=9/2)$ kernel, as shown in the top panels of Fig.~\ref{Fig3}. Within the standard quintessence framework in general relativity, this behavior is unphysical, as it implies $\dot{\phi}^2 < 0$ and therefore violates the null energy condition \cite{copeland2006dynamics}. Although the median reconstruction of $\tau(z)$ becomes negative at intermediate and high redshift, the corresponding $1\sigma$ region remains consistent with positive values. At high redshift, the uncertainties become large, suggesting that the apparent phantom-like behavior is more likely a consequence of the limited constraining power of the data in this regime than a robust physical effect.

The corresponding potentials $U(z)$, together with their $1\sigma$ and $2\sigma$ confidence intervals for both kernels, are shown in the lower panels of Fig.~\ref{Fig3}. For comparison, we include two theoretical benchmark models: the Power Law (PL) model, $U_{\mathrm{PL}}(z)$, and the exponential (EXP) model, $U_{\mathrm{EXP}}(z)$. Following the analyses of Refs.~\cite{cao2020cosmological} and \cite{bhattacharya2024cosmological}, we adopt $\alpha = 0.150$ for the PL model and $\lambda = 0.48$ for the EXP model. The reconstructed potential, $U(z)$, is compatible with the PL model within the $1\sigma$ confidence interval from $z \approx 0.38$ up to $z \approx 2.05$ for both kernels. The exponential model $U_{\mathrm{EXP}}(z)$ agrees with the reconstruction at the $1\sigma$ level for redshifts $z \leq 1.8$, beyond which it departs from the $1\sigma$ band, while remaining within the $2\sigma$ confidence interval over the full reconstructed redshift range for both kernels. 

Using the reconstructed $D_M(z)$, and its derivatives in Eqs.~(\ref{U_D}) and (\ref{tau_D}), we also reconstruct $U(z)$ and $\tau(z)$ from the Pantheon+ dataset. The results are shown in Fig.~ \ref{Fig4}. In the top panels, the reconstructed kinetic term, $\tau(z)$ remains positive over the entire reconstructed redshift range for both kernels, although the associated uncertainties increase at high redshift.  This feature can be attributed to the sharp decline in Pantheon+ data density beyond $z \approx 0.8$, combined with the amplification of uncertainties inherent in the reconstruction of second order derivatives.

The corresponding potentials are shown in the lower panels of Fig.~\ref{Fig4}. 
The $U_{\mathrm{PL}}(z)$ curve is compatible with the reconstructed potential within the $1\sigma$ confidence interval over approximately $68\%$ of the reconstructed redshift range for the squared exponential kernel, and over approximately $70\%$ for the Matern ($\nu = 9/2$) kernel. 
The exponential model $U_{\mathrm{EXP}}(z)$ is consistent with the reconstruction at the $1\sigma$ level over approximately $48\%$ of the redshift range for the squared exponential kernel, and over approximately $61\%$ for the Matern ($\nu = 9/2$) kernel.  Both benchmark models remain entirely within the $2\sigma$ confidence intervals over the full reconstructed redshift range.

Finally, we note that when using Pantheon+ data, small differences arise between the squared exponential and Matern ($\nu = 9/2$) kernels. This contrasts with the Hubble data reconstructions, which are largely insensitive to the kernel choice. This behavior originates from the requirement to reconstruct second order derivatives when using Pantheon+ data. While the squared exponential kernel is infinitely differentiable, the Matern ($\nu = 9/2$) kernel, which is four times mean-square differentiable, yields mild differences in higher order derivative reconstructions.

\begin{figure}[!t]
\hspace{-0.08 cm}
\begin{minipage}[t]{0.52\textwidth}
\includegraphics[width=\linewidth]{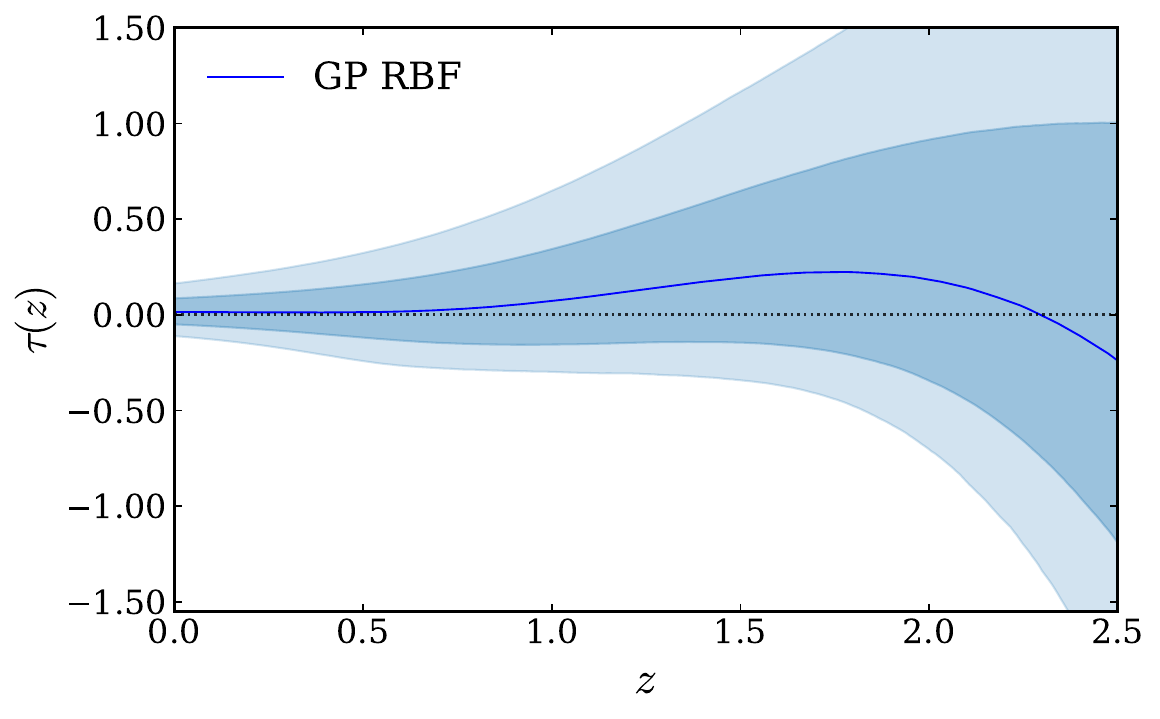}
\end{minipage}%
\hspace{-0.22 cm}
\begin{minipage}[t]{0.52\textwidth}
\includegraphics[width=\linewidth]{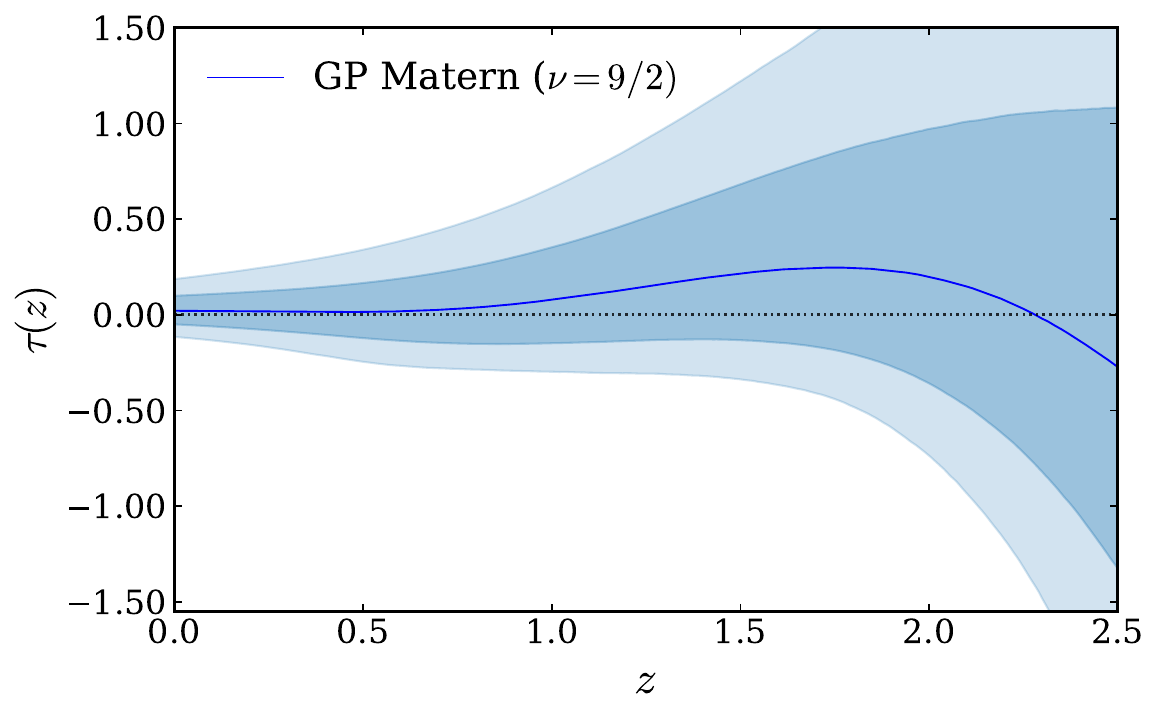}
\end{minipage}

\begin{minipage}[t]{0.52\textwidth}
\includegraphics[width=\linewidth]{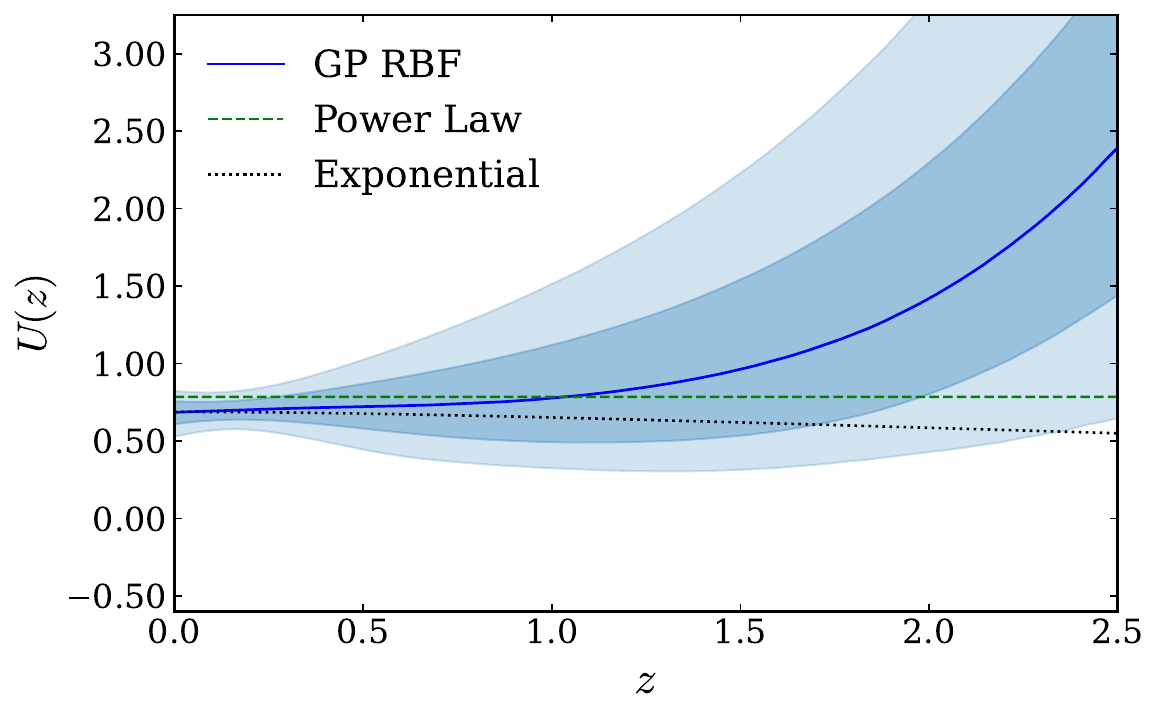}
\end{minipage}%
\hspace{-0.1 cm}
\begin{minipage}[t]{0.52\textwidth}
\includegraphics[width=\linewidth]{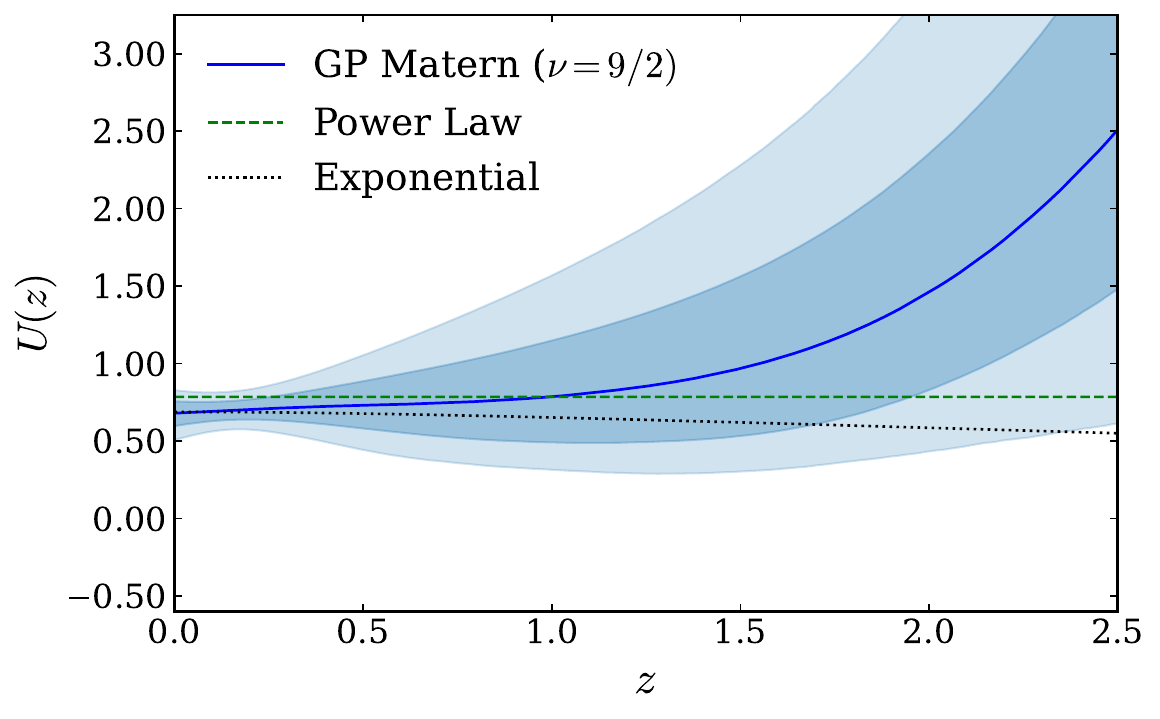}
\end{minipage}
    
\caption{Reconstruction of the dimensionless kinetic energy $\tau(z)$ (top panels) and the scalar field potential $U(z)$ (bottom panels) from Hubble parameter measurements, assuming a Large prior on $(\Omega_m,\Omega_k)$. The left panels correspond to the squared exponential (RBF) kernel, while the right panels correspond to the Matern ($\nu=9/2$) kernel. The blue and light blue shaded regions denote the 1$\sigma$ and 2$\sigma$ confidence intervals, respectively. The green dashed and black dotted curves represent the power law and exponential potentials, respectively.}

\label{Fig5}
\end{figure}\

\subsection*{B. Reconstruction of $U(z)$ and $\tau(z)$ Using the Large Prior}\ 

We follow the same procedure as in Sec.~\hyperref[secA]{A} to reconstruct $U(z)$ and $\tau(z)$ using the Large prior for the matter and curvature density parameters. The top panels of Fig.~\ref{Fig5} show the reconstruction of $\tau(z)$ from Hubble data. In this case, the median kinetic energy, $\tau(z)$, crosses zero and becomes negative at $z \approx 2.28$ for both kernels. The lower panels of Fig.~\ref{Fig5} display the reconstruction of $U(z)$. The $U_{\mathrm{PL}}(z)$ and $U_{\mathrm{EXP}}(z)$ curves are compatible 
with the reconstructed potential within the $1\sigma$ confidence interval 
over approximately $67\%$ of the analyzed redshift range, for both kernels.

\begin{figure}[!t]
\begin{minipage}[t]{0.5\textwidth}
\includegraphics[width=\linewidth]{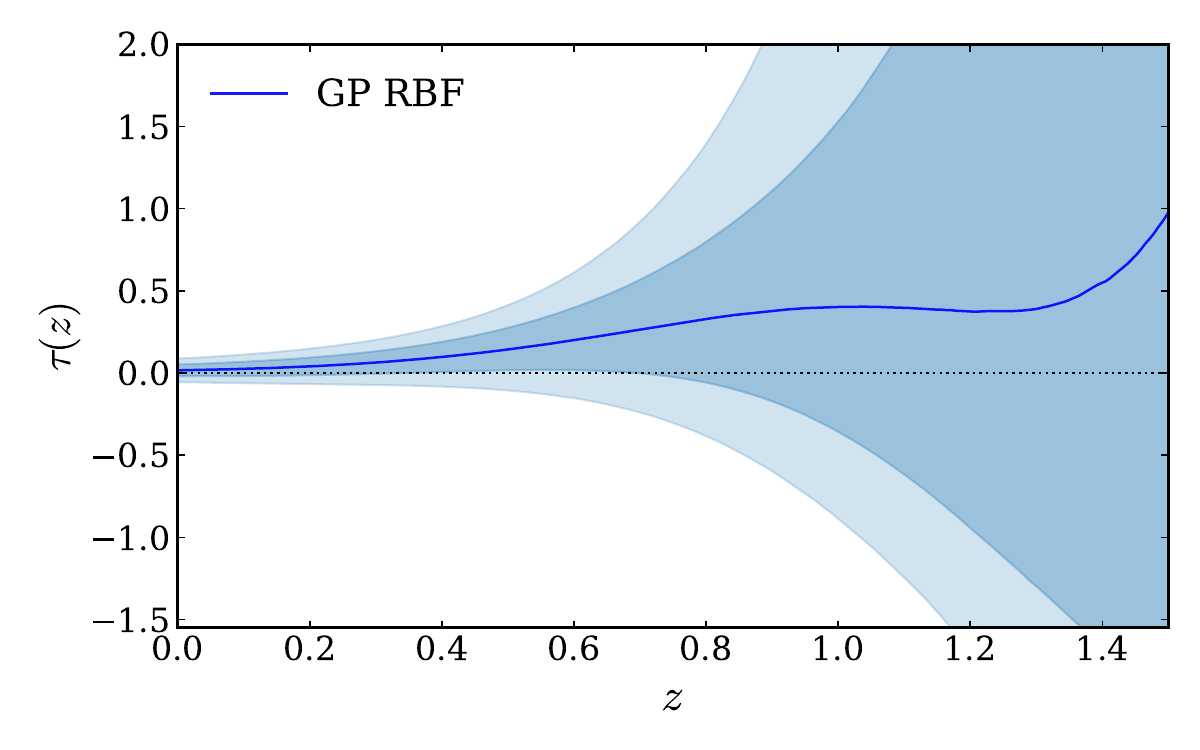}
\end{minipage}%
\begin{minipage}[t]{0.5\textwidth}
\includegraphics[width=\linewidth]{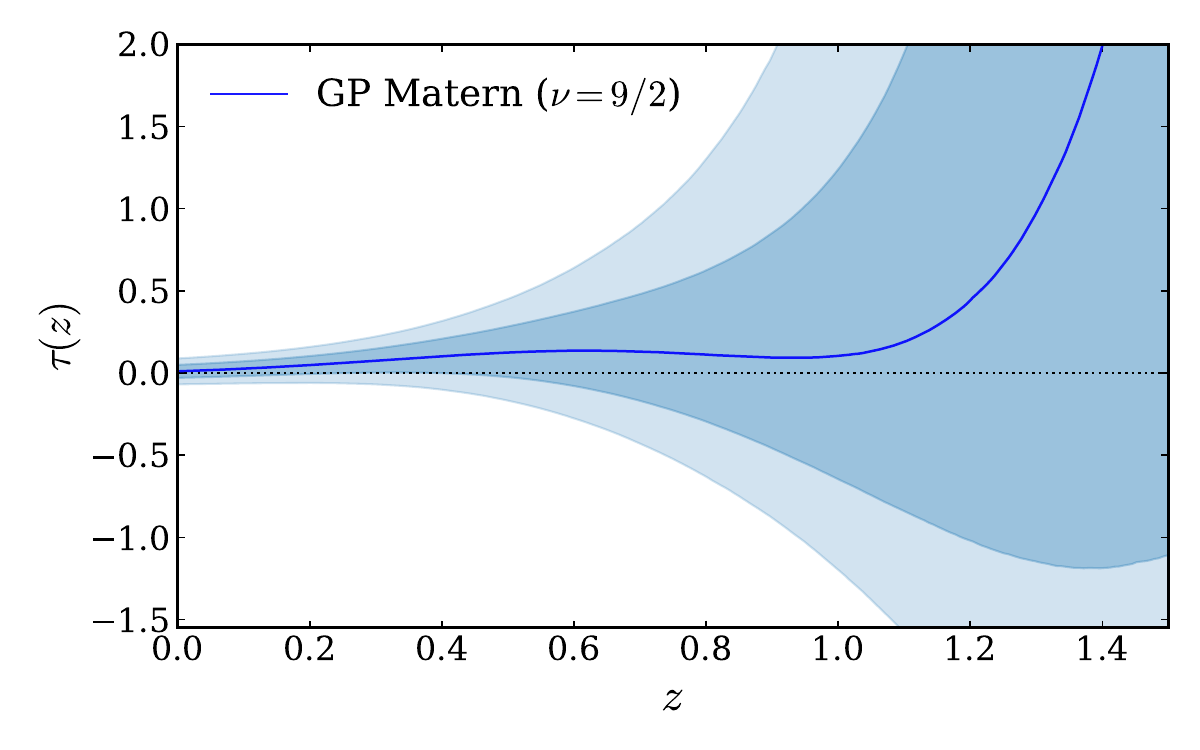}
\end{minipage}

\begin{minipage}[t]{0.54\textwidth}
\raisebox{0.53cm}{\includegraphics[width=\linewidth]{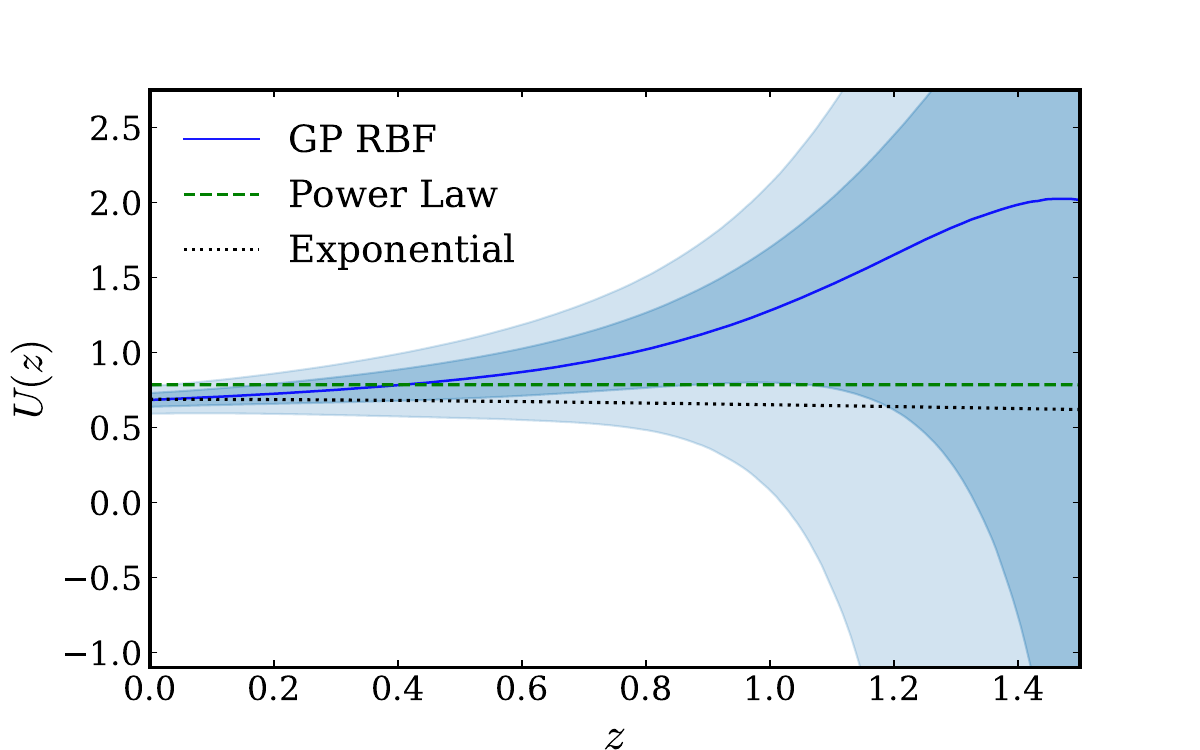}}
\end{minipage}%
\hspace{-0.5 cm}
\begin{minipage}[t]{0.54\textwidth}
\raisebox{0.53cm}{\includegraphics[width=\linewidth]{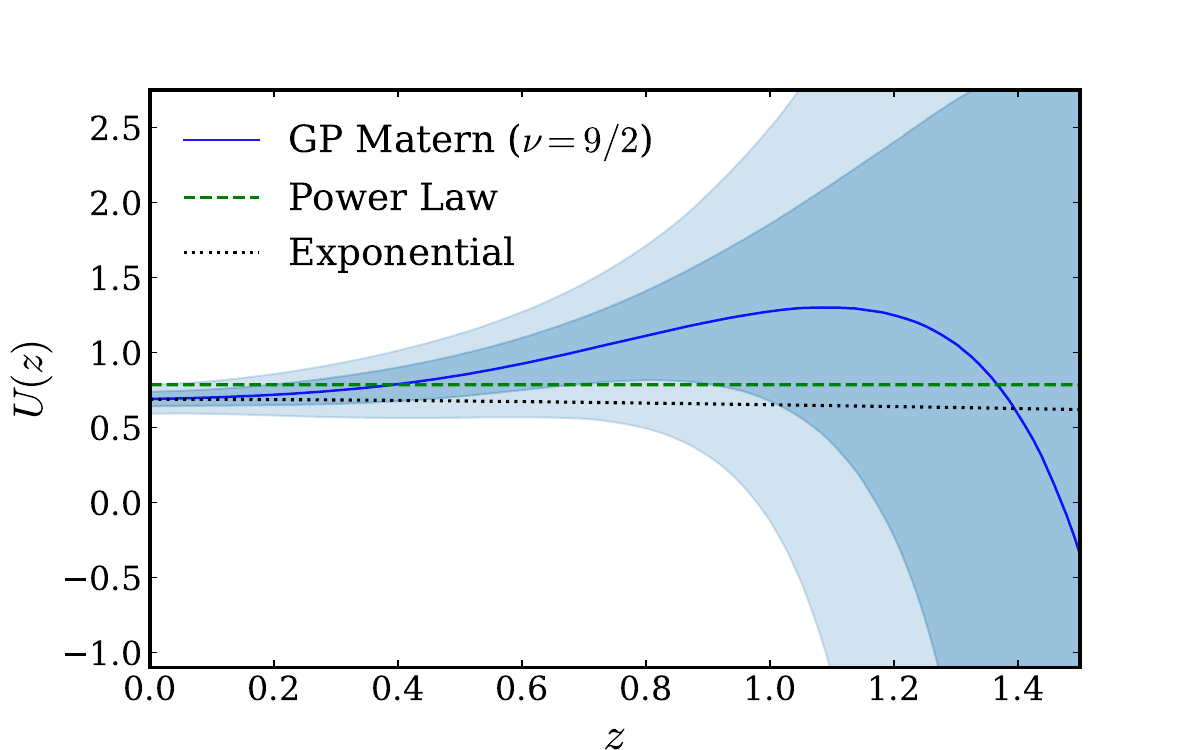}}
\end{minipage}
    
\caption{Reconstruction of the scalar-field kinetic term (top panels) and potential (bottom panels) from Pantheon+ data, assuming a Large prior on $(\Omega_m,\Omega_k)$. The left panels correspond to the squared exponential (RBF) kernel, while the right panels correspond to the Matern ($\nu=9/2$) kernel. The blue and light blue shaded areas indicate the 1$\sigma$ and 2$\sigma$ confidence intervals, respectively. The green dashed and black dotted curves represent the power law and exponential potentials, respectively.}

\label{Fig6}
\end{figure}

Fig.~\ref{Fig6} presents the reconstruction based on the Pantheon+ dataset. The median $\tau(z)$ remains positive over the entire reconstructed 
redshift range for both kernels. As expected, the confidence intervals widen at higher redshift due to data sparsity and the amplification of uncertainties associated with higher-order derivatives. The lower panels of Fig.~\ref{Fig6} show the reconstructed potential $U(z)$. Using the squared exponential kernel, the $U_{PL}(z)$ and $U_{EXP}(z)$ curves are compatible within the $1\sigma$ confidence interval over approximately $82\%$ and $61\%$ of the redshift range, respectively. With the Matern ($\nu = 9/2$) kernel,  the $U_{PL}(z)$ curve is compatible within $1\sigma$ over approximately $79\%$, while the $U_{EXP}(z)$ curve remains compatible at the $1\sigma$ confidence interval over $70\%$  of the redshift range.

\

\begin{figure}[!t]
\begin{minipage}[t]{0.52\textwidth}
\includegraphics[width=\linewidth]{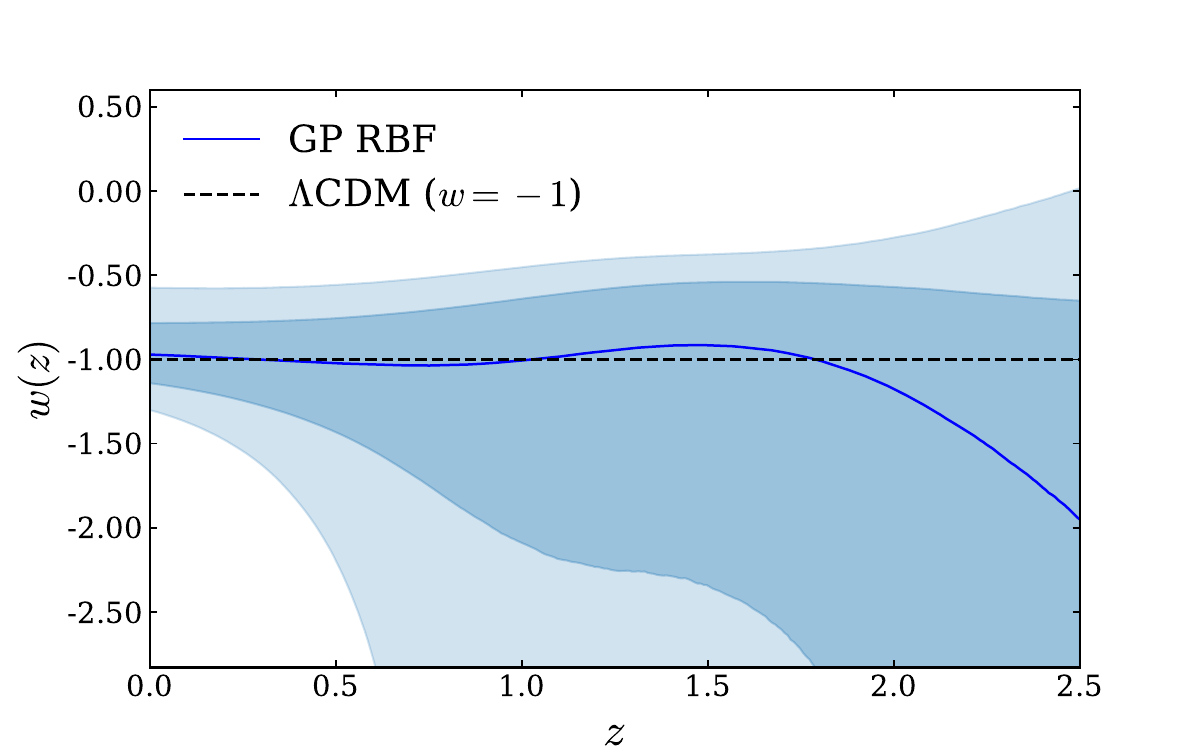}
\end{minipage}%
\begin{minipage}[t]{0.52\textwidth}
\includegraphics[width=\linewidth]{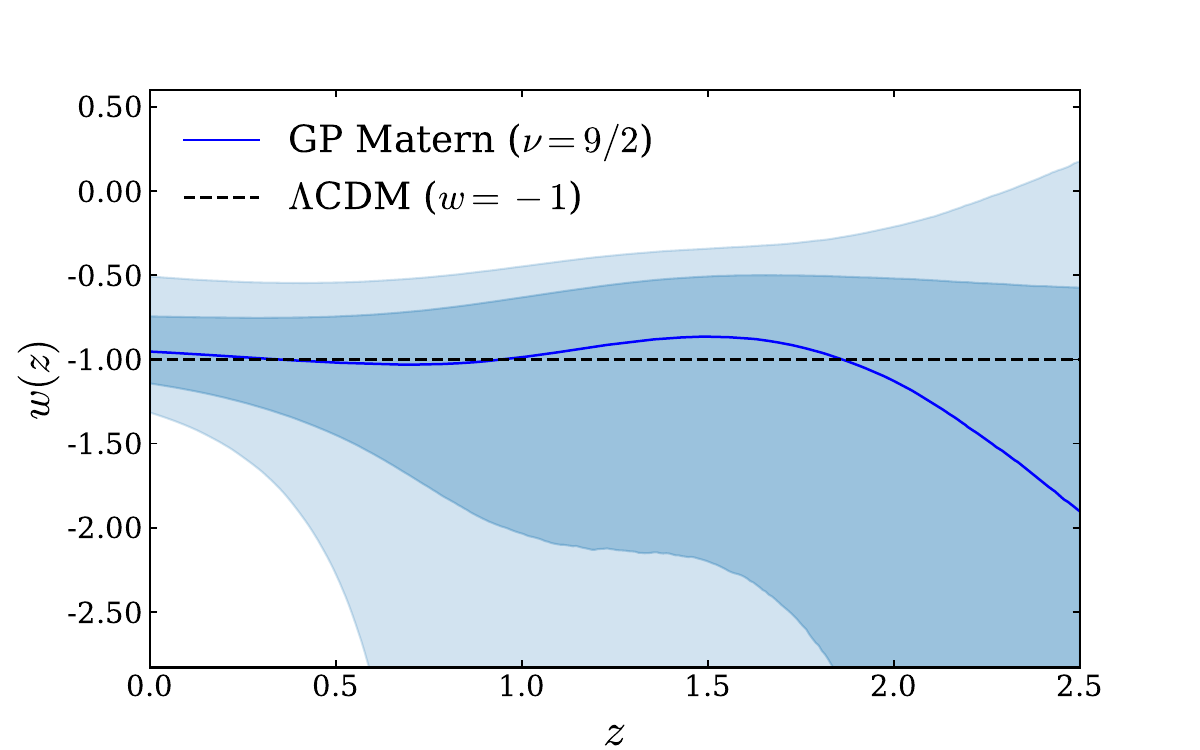}
\end{minipage}

\begin{minipage}[t]{0.52\textwidth}
\includegraphics[width=\linewidth]{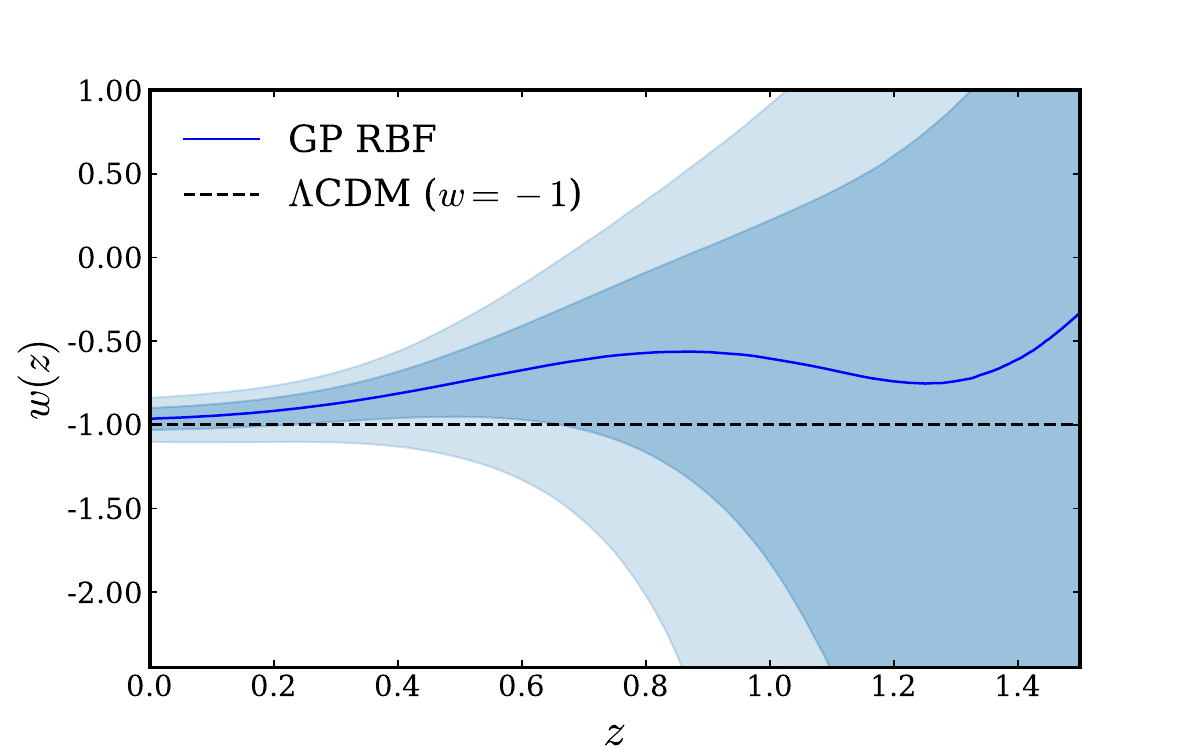}
\end{minipage}%
\begin{minipage}[t]{0.52\textwidth}
\includegraphics[width=\linewidth]{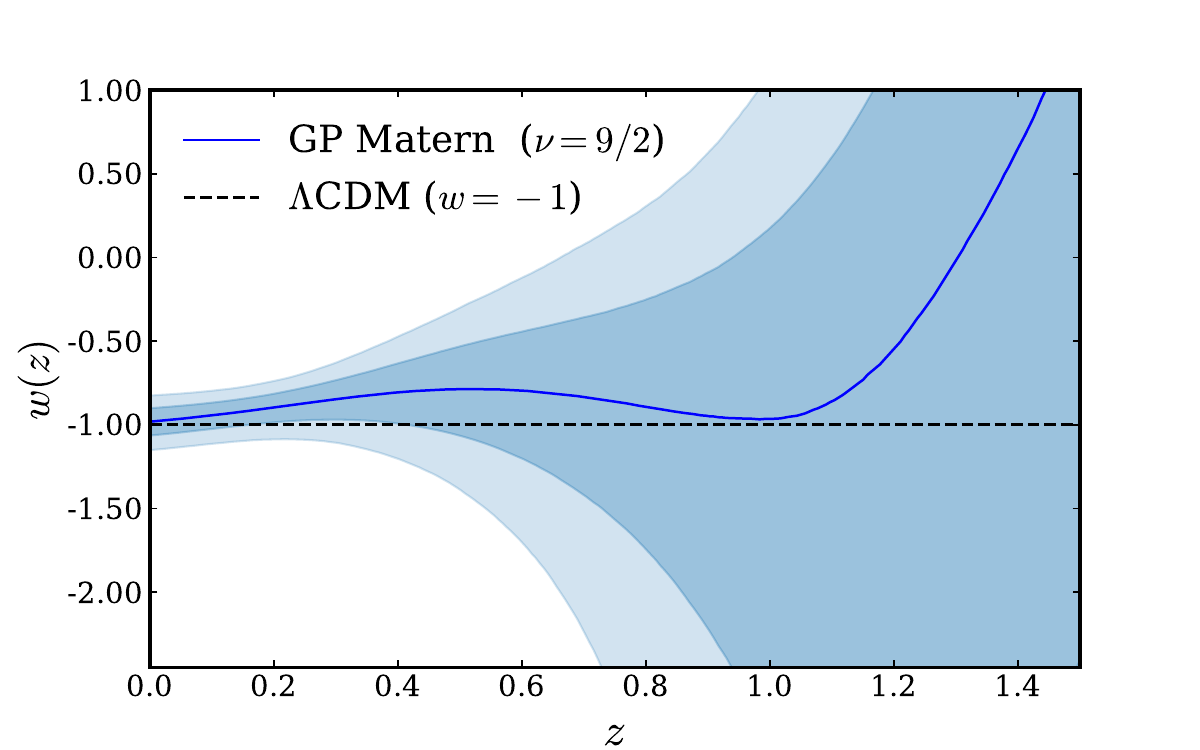}

\end{minipage}

\caption{Reconstruction of the dark energy equation of state parameter $w(z)$ from $H(z)$ measurements (top panels) and from the Pantheon+ data (bottom panels), assuming a Planck 3$\sigma$ prior on $(\Omega_m,\Omega_k)$. The left panels correspond to the squared exponential (RBF) kernel, while the right panels correspond to the Matern ($\nu=9/2$) kernel. The blue and light blue shaded regions denote the 1$\sigma$ and 2$\sigma$ confidence intervals, respectively. The black dashed line indicates  the $\Lambda$CDM model.}
\label{fig7}
\end{figure}

\begin{figure}[!t]
\begin{minipage}[t]{0.52\textwidth}
\includegraphics[width=\linewidth]{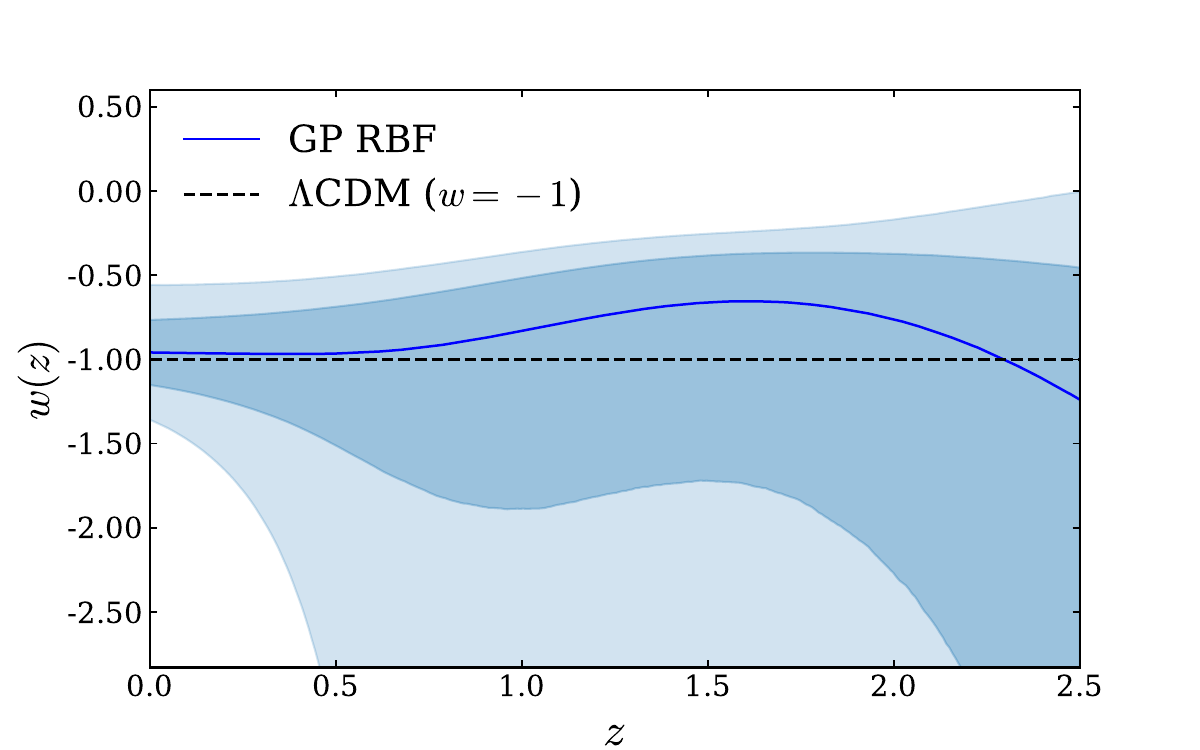}
\end{minipage}%
\hspace{-0.2 cm}
\begin{minipage}[t]{0.52\textwidth}
\includegraphics[width=\linewidth]{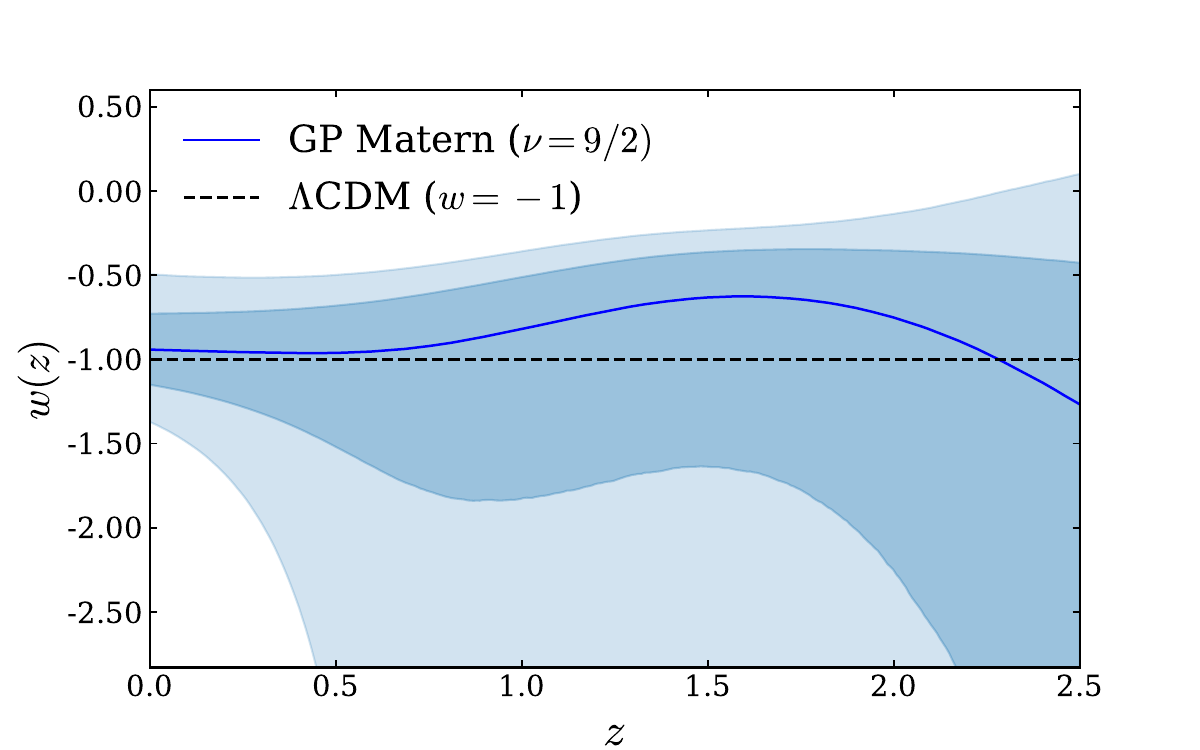}
\end{minipage}

\begin{minipage}[t]{0.52\textwidth}
\includegraphics[width=\linewidth]{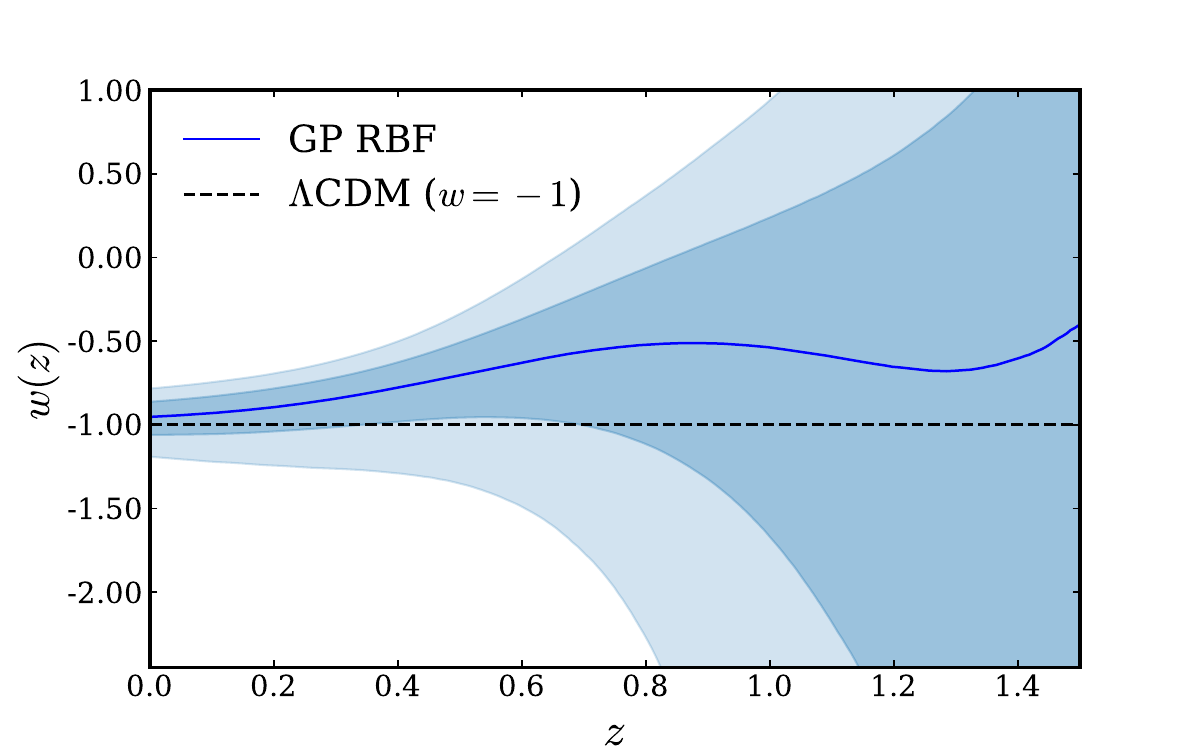}
\end{minipage}%
\hspace{-0.2 cm}
\begin{minipage}[t]{0.52\textwidth}
\includegraphics[width=\linewidth]{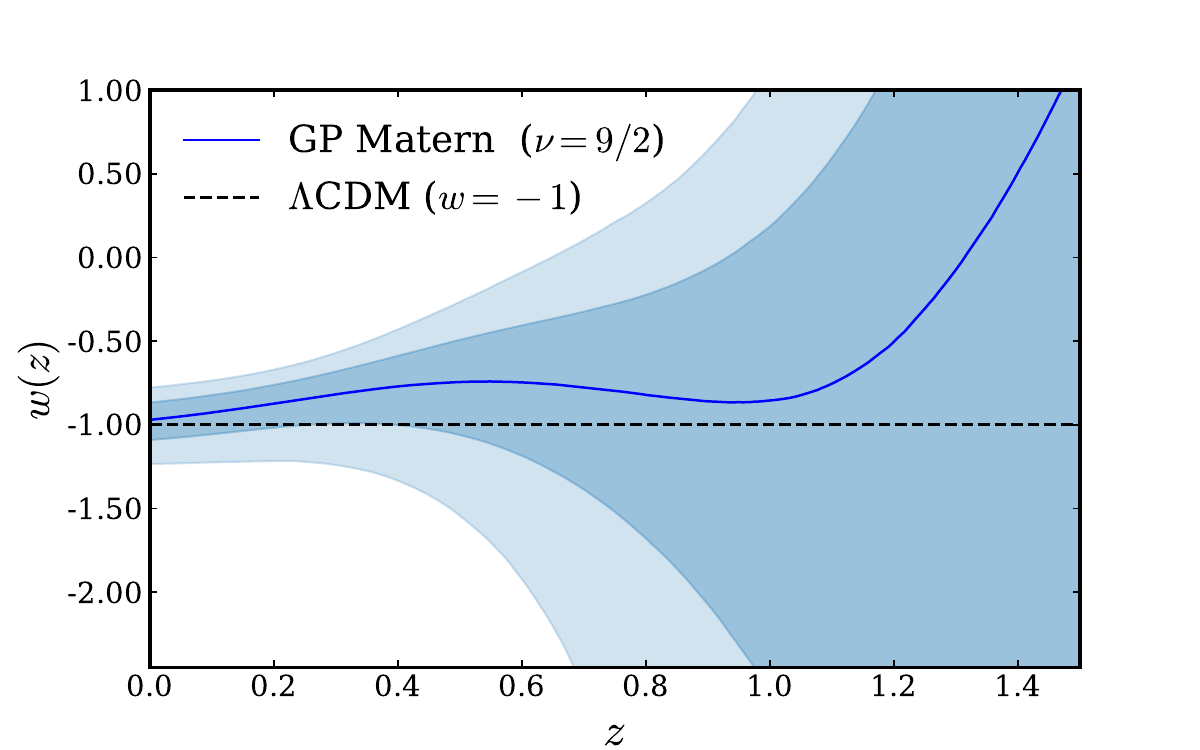}
\end{minipage}
\caption{Reconstruction of the dark energy equation of state parameter $w(z)$ from $H(z)$ measurements (top panels) and from the Pantheon+ data (bottom panels), assuming a large prior on $(\Omega_m,\Omega_k)$. The left panels correspond to the squared exponential (RBF) kernel, while the right panels correspond to the Matern ($\nu=9/2$) kernel. The blue and light blue shaded regions denote the 1$\sigma$ and 2$\sigma$ confidence intervals, respectively. The black dashed line indicates  the $\Lambda$CDM model.}
\label{fig8}
\end{figure}

The potential and kinetic reconstructions presented above allow us to 
directly infer the equation-of-state parameter of the quintessence field, defined as
\begin{equation}
w(z) = \frac{p_\phi}{\rho_\phi} =\frac{\tau(z) - U(z)}{\tau(z) + U(z)},
\end{equation}
and is displayed in Figs.~\ref{fig7} and \ref{fig8}. For a canonical scalar field 
with a non-negative kinetic term, the equation-of-state parameter satisfies 
$w \geq -1$ by construction, with $w = -1$ recovered only in the 
potential dominated limit $\tau \rightarrow 0$.

Considering the $H(z)$ measurements with the Planck-based prior, the reconstructed value at the present epoch is $w(0) = -0.971^{+0.186}_{-0.169}$ for the squared exponential  kernel 
and $w(0) = -0.953^{+0.208}_{-0.189}$ for the Matern ($\nu = 9/2$) kernel. 
In both cases, the median reconstructed value lies above $-1$, as the reconstructed kinetic term, $\tau(z)$, remains non-negligible at $z = 0$. Using the Pantheon+ dataset with the same prior, the reconstruction 
yields $w(0) = -0.963^{+0.064}_{-0.067}$ for the squared exponential kernel and 
$w(0) = -0.981^{+0.080}_{-0.082}$ for the Matern ($\nu = 9/2$) kernel. 
These values are also close to $-1$, suggesting that the scalar field 
is closer to its potential dominated regime at the present epoch. In all four cases, the cosmological constant 
$w = -1$ remains consistent with the reconstruction within the $1\sigma$ 
confidence interval.

Adopting the large prior, the reconstruction based on the Hubble measurements gives 
$w(0) = -0.958^{+0.192}_{-0.193}$ for the squared exponential kernel and 
$w(0) = -0.941^{+0.213}_{-0.209}$ for the Matern ($\nu = 9/2$) kernel. For the Pantheon+ dataset 
with the large prior, we obtain $w(0) = -0.952^{+0.090}_{-0.107}$ for the squared exponential 
kernel and $w(0) = -0.970^{+0.102}_{-0.120}$ for the Matern ($\nu = 9/2$) 
kernel.

For the $H(z)$ reconstruction with the Planck-based prior, the median reconstructed equation-of-state parameter crosses $w=-1$ at intermediate and high redshift for both kernels. For the larger prior, the corresponding crossing occurs only at high redshift. This difference indicates that the reconstructed $w(z)$ is sensitive to the adopted values of the matter and curvature density parameters, which motivates examining more than one prior choice.

In contrast, the Pantheon+ reconstruction remains above $w=-1$ over the reconstructed redshift range. At low redshift, this reconstruction appears better constrained, likely owing to the much larger size of the Pantheon+ sample compared with the current $H(z)$ dataset.
Across all configurations, the reconstructed $w(z)$ remains consistent with the quintessence region at low redshift, where the data provide the strongest constraints, while also remaining compatible with $w=-1$ within the corresponding $1\sigma$ confidence interval.

\color{black}

As discussed in Ref.~\cite{dinda2025physical}, the apparent preference for quintom-like behavior in recent cosmological data, namely, a crossing from phantom to quintessence regimes, depends sensitively on the adopted parametrization of the dark energy equation of state, such as the CPL form. In particular, thawing quintessence models with nonzero spatial curvature can reproduce the observational data without invoking $w < -1$, whereas phenomenological parametrizations in a flat Universe may artificially suggest a phantom crossing. This highlights that, although realistic scalar field models forbid $w < -1$, phantom like behavior can appear in parametrizations not grounded in a physical model.

\section{Conclusions}\label{sec5}\  

In this work, we have performed a model independent reconstruction of the quintessence scalar field potential, $U(z)$, and its associated kinetic term, $\tau(z)$, using Gaussian Processes and current cosmological observations. The analysis employed two complementary datasets, namely recent measurements of the Hubble parameter and the Pantheon+ Type Ia supernova compilation, under two distinct assumptions for the cosmological density parameters: a Planck informed prior and a large prior. This strategy allowed us to assess the robustness of the reconstruction with respect to both prior choices and kernel dependence.

We reconstructed the Hubble parameter, $H(z)$, and its derivative from Hubble parameter measurements, as well as the dimensionless transverse comoving distance, $D_M(z)$, and its derivatives from Pantheon+ data. Two different covariance functions were employed: the squared exponential kernel and the Matern ($\nu=9/2$) kernel. This approach allows us to test the sensitivity of the inferred scalar field dynamics to the assumed kernel structure.

For each dataset, we reconstructed the dimensionless scalar field potential, $U(z)$, and its associated kinetic term, $\tau(z)$. To assess the impact of background cosmological assumptions, we considered two choices for the matter density and spatial curvature parameters: (i) a Planck-based prior within $3\sigma$ confidence level, motivated by the fact that these constraints assume the $\Lambda$CDM model rather than a dynamical scalar field dark energy cosmologies, and (ii) a large prior that allows greater flexibility. When using Pantheon+ data, small differences arise between the squared exponential and Matern ($\nu=9/2$) kernels, due to the need to reconstruct second order derivatives. While the squared exponential kernel is infinitely differentiable, the Matern ($\nu=9/2$) kernel is only four times mean-square differentiable, which may lead to mild differences in the reconstruction of higher order derivatives. In contrast, the Hubble data reconstructions are largely insensitive to the choice of kernel.Furthermore, the reconstruction exhibits some dependence on the matter and curvature density parameters, motivating the examination of multiple prior choices.

We further compared the reconstructed potential with two theoretically motivated benchmark models: a power law potential, $U_{\rm PL}(z)$, and an exponential potential, $U_{\rm EXP}(z)$. For both Hubble and Pantheon+ data, and for both prior choices and kernels, these benchmark models remain consistent with the reconstructed potential within the $1\sigma$ confidence interval for most of the redshift range explored, and within the $2\sigma$ confidence interval over the remaining redshift range.

The reconstructed quintessence potential is obtained in a model independent manner, allowing the data to guide the results directly. Both Hubble and Pantheon+ datasets favor quintessence like behavior at low redshift. Although the median kinetic energy, $\tau(z)$, becomes negative at high redshift, which is unphysical within the standard quintessence framework, the associated uncertainties are large, preventing a definitive conclusion regarding phantom like behavior. 
 This suggests the negative $\tau(z)$ is most likely an artifact of limited high redshift data rather than a genuine physical effect.

We have reconstructed the equation-of-state parameter $w(z)$, and find that its median remains in the quintessence regime but that the reconstruction is consistent with a variety of scenarios, such as a cosmological constant, phantom dark energy, and phantom crossing dark energy. At the present epoch, the confidence regions for Pantheon+ data are smaller than those for the $H(z)$ measurements, indicating a higher precision in the reconstruction. At higher redshifts, the uncertainties increase due to the sparser coverage of the data.

Overall, our results support the view that current cosmological data are consistent with standard quintessence dynamics at late times. Apparent indications of phantom or quintom like behavior in non parametric reconstructions should be interpreted with caution, as they may reflect the limited redshift coverage and precision of current datasets. Future surveys with broader redshift coverage and higher precision will be essential to further constrain the scalar field dynamics of dark energy and to discriminate between physically motivated models in a fully model independent framework.

\section*{Acknowledgements}\ 

 R. E. O. is grateful for support from the \enquote{PhD-Associate Scholarship – PASS} grant (number 29 UMP2023) provided by the National Center for Scientific and Technical Research in Morocco.

\end{document}